\documentclass[%
 reprint,
superscriptaddress,
 amsmath,amssymb,
 aps,
]{revtex4-2}

\usepackage{graphicx}
\usepackage{dcolumn}
\usepackage{bm}
\usepackage{hyperref}
\usepackage{cleveref}
\usepackage{float}

\begin{document}

\newcommand{\caseabrokenqubithex}{https://algassert.com/crumble\#circuit=Q(1,5)0;Q(2,4)1;Q(2,5)2;Q(2,6)3;Q(3,3)4;Q(3,4)5;Q(3,5)6;Q(3,6)7;Q(3,7)8;Q(4,2)9;Q(4,3)10;Q(4,4)11;Q(4,5)12;Q(4,6)13;Q(4,7)14;Q(4,8)15;Q(5,1)16;Q(5,2)17;Q(5,3)18;Q(5,4)19;Q(5,6)20;Q(5,7)21;Q(5,8)22;Q(5,9)23;Q(6,1)24;Q(6,2)25;Q(6,3)26;Q(6,4)27;Q(6,5)28;Q(6,6)29;Q(6,7)30;Q(6,8)31;Q(6,9)32;Q(7,2)33;Q(7,3)34;Q(7,4)35;Q(7,5)36;Q(7,6)37;Q(7,7)38;Q(7,8)39;Q(8,3)40;Q(8,4)41;Q(8,5)42;Q(8,6)43;Q(8,7)44;Q(9,4)45;Q(9,5)46;Q(9,6)47;R_47_14_10_30_26_44_39_19_32_2_12_28_42;RX_16_9_25_4_18_34_1_11_27_41_0_6_36_46_3_13_29_43_8_21_38_15_31_23_24_37_17_33_22_40_5_35_45_7_20;TICK;CX_20_13_21_30_35_27_33_25_37_29_6_12_22_15_4_10_18_26_7_3_5_1_38_44_45_41_23_32_17_9_8_14_36_42_0_2;TICK;CX_22_23_25_26_41_42_33_34_20_21_45_46_29_30_31_32_1_2_9_10_5_6_27_28_37_38_17_18_7_8_11_12_35_36_43_44_13_14;TICK;CX_2_3_6_7_18_19_14_15_28_29_21_22_12_13_30_31_26_27_34_35_4_5_36_37_42_43_16_17_38_39_24_25_46_47_40_41_10_11;TICK;CX_31_39_1_5_12_6_26_18_24_16_14_8_29_37_11_19_30_21_40_34_3_7_10_4_42_36_15_22_2_0_43_47_9_17_13_20_27_35;TICK;M_37_17_47_44_22_5_39_19_32_35_7_20;MX_24_14_33_10_30_26_40_45_2_12_42;DT(6,1,0)rec[-11];DT(4,7,0)rec[-10];DT(7,2,0)rec[-9];DT(4,3,0)rec[-8];DT(6,7,0)rec[-7];DT(8,3,0)rec[-5];DT(9,4,0)rec[-4];DT(2,5,0)rec[-3];DT(4,5,0)rec[-2];DT(8,5,0)rec[-1];TICK;R_37_17_47_44_22_5_39_19_32_35_7_20;RX_24_14_33_10_30_26_40_45_2_12_42;TICK;CX_31_39_1_5_12_6_26_18_24_16_14_8_29_37_11_19_30_21_40_34_3_7_10_4_42_36_15_22_2_0_43_47_9_17_13_20_27_35;TICK;CX_2_3_6_7_18_19_14_15_28_29_21_22_12_13_30_31_26_27_34_35_4_5_36_37_42_43_16_17_38_39_24_25_46_47_40_41_10_11;TICK;CX_22_23_25_26_41_42_33_34_20_21_45_46_29_30_31_32_1_2_9_10_5_6_27_28_37_38_17_18_7_8_11_12_35_36_43_44_13_14;TICK;CX_33_25_6_12_18_26_38_44_45_41_8_14_19_11_37_29_22_15_7_3_23_32_4_10_17_9_0_2_20_13_21_30_35_27_5_1_36_42;TICK;M_47_14_10_30_26_44_39_32_2_12_42;MX_24_37_17_33_22_40_5_19_35_45_7_20;DT(9,6,1)rec[-23]_rec[-44];DT(3,6,1)rec[-22]_rec[-36];DT(4,3,1)rec[-21];DT(5,6,1)rec[-20]_rec[-35];DT(5,2,1)rec[-19]_rec[-45];DT(8,7,1)rec[-18]_rec[-43]_rec[-46];DT(7,8,1)rec[-17]_rec[-40];DT(6,9,1)rec[-16]_rec[-38]_rec[-42];DT(2,5,1)rec[-15];DT(3,4,1)rec[-14]_rec[-41];DT(7,4,1)rec[-13]_rec[-37];DT(6,1,1)rec[-12]_rec[-34];DT(6,7,1)rec[-11]_rec[-30];DT(4.5,2.5,1)rec[-10]_rec[-31];DT(6,3,1)rec[-9]_rec[-29]_rec[-32];DT(5,8,1)rec[-8];DT(8,3,1)rec[-7]_rec[-28];DT(2.5,4.5,1)rec[-6]_rec[-26];DT(7,4,2)rec[-4];DT(8,5,1)rec[-3]_rec[-24]_rec[-27];DT(3,6,2)rec[-2];DT(4,7,1)rec[-1]_rec[-33];TICK;R_47_14_10_30_26_44_39_32_2_12_42;RX_24_37_17_33_22_40_5_19_35_45_7_20;TICK;CX_33_25_6_12_18_26_38_44_45_41_8_14_19_11_37_29_22_15_7_3_23_32_4_10_17_9_0_2_20_13_21_30_35_27_5_1_36_42;TICK;CX_22_23_25_26_41_42_33_34_20_21_45_46_29_30_31_32_1_2_9_10_5_6_27_28_37_38_17_18_7_8_11_12_35_36_43_44_13_14;TICK;CX_2_3_6_7_18_19_14_15_28_29_21_22_12_13_30_31_26_27_34_35_4_5_36_37_42_43_16_17_38_39_24_25_46_47_40_41_10_11;TICK;CX_2_0_40_34_31_39_3_7_27_35_1_5_29_37_43_47_9_17_12_6_24_16_10_4_26_18_42_36_15_22_11_19_30_21_14_8;TICK;M_37_17_47_44_22_5_39_19_32_35_7;MX_24_14_33_10_30_26_40_45_2_12_28_42_20;DT(7,6,3)rec[-24];DT(5,2,3)rec[-23];DT(8,5,3)rec[-22]_rec[-37]_rec[-47];DT(8,7,3)rec[-21]_rec[-42];DT(4,7,3)rec[-20]_rec[-46];DT(3,4,3)rec[-19];DT(7,8,3)rec[-18]_rec[-41]_rec[-44];DT(4,3,3)rec[-17]_rec[-45];DT(6,9,3)rec[-16]_rec[-40];DT(6,3,3)rec[-15]_rec[-43];DT(2,5,3)rec[-14]_rec[-39];DT(6,1,3)rec[-13]_rec[-34]_rec[-36];DT(4,7,4)rec[-12];DT(7,2,3)rec[-11]_rec[-33];DT(3.5,3.5,3)rec[-10]_rec[-30];DT(5,8,3)rec[-9]_rec[-32];DT(5,4,3)rec[-8]_rec[-29];DT(7,4,3)rec[-7]_rec[-28]_rec[-31];DT(9,4,3)rec[-6]_rec[-27];DT(2,5,4)rec[-5];DT(3,6,3)rec[-4]_rec[-26];DT(7.5,5.5,3)rec[-2]_rec[-35];DT(4,5,3)rec[-1]_rec[-3]_rec[-25]_rec[-29]_rec[-49]_rec[-53];TICK;R_37_17_47_44_22_5_39_19_32_35_7;RX_24_14_33_10_30_26_40_45_2_12_28_42_20;TICK;CX_2_0_40_34_31_39_3_7_27_35_1_5_29_37_43_47_9_17_12_6_24_16_10_4_26_18_42_36_15_22_11_19_30_21_14_8;TICK;CX_2_3_6_7_18_19_14_15_28_29_21_22_12_13_30_31_26_27_34_35_4_5_36_37_42_43_16_17_38_39_24_25_46_47_40_41_10_11;TICK;CX_22_23_25_26_41_42_33_34_20_21_45_46_29_30_31_32_1_2_9_10_5_6_27_28_37_38_17_18_7_8_11_12_35_36_43_44_13_14;TICK;CX_20_13_21_30_35_27_33_25_37_29_6_12_22_15_4_10_18_26_7_3_5_1_38_44_45_41_23_32_17_9_8_14_36_42_0_2;TICK;M_47_14_10_30_26_44_39_19_32_2_12_28_42;MX_24_37_17_33_22_40_5_35_45_7_20;DT(9,6,5)rec[-24]_rec[-46];DT(3,6,5)rec[-23]_rec[-38];DT(4,3,5)rec[-22];DT(5,2,5)rec[-20]_rec[-47];DT(8,7,5)rec[-19]_rec[-45]_rec[-48];DT(7,8,5)rec[-18]_rec[-42];DT(6,9,5)rec[-16]_rec[-40]_rec[-44];DT(2,5,5)rec[-15];DT(3,4,5)rec[-14]_rec[-43];DT(7,4,5)rec[-12]_rec[-39];DT(6,1,5)rec[-11]_rec[-37];DT(6,7,5)rec[-10]_rec[-33];DT(4.5,2.5,5)rec[-9]_rec[-34];DT(6,3,5)rec[-8]_rec[-32]_rec[-35];DT(5,8,5)rec[-7];DT(8,3,5)rec[-6]_rec[-31];DT(2.5,4.5,5)rec[-5]_rec[-29];DT(6,5,5)rec[-4]_rec[-27];DT(8,5,5)rec[-3]_rec[-26]_rec[-30];DT(3,6,6)rec[-2];DT(5,6,5)rec[-1]_rec[-25]_rec[-36];TICK;R_47_14_10_30_26_44_39_19_32_2_12_28_42;RX_24_37_17_33_22_40_5_35_45_7_20;TICK;CX_20_13_21_30_35_27_33_25_37_29_6_12_22_15_4_10_18_26_7_3_5_1_38_44_45_41_23_32_17_9_8_14_36_42_0_2;TICK;CX_22_23_25_26_41_42_33_34_20_21_45_46_29_30_31_32_1_2_9_10_5_6_27_28_37_38_17_18_7_8_11_12_35_36_43_44_13_14;TICK;CX_2_3_6_7_18_19_14_15_28_29_21_22_12_13_30_31_26_27_34_35_4_5_36_37_42_43_16_17_38_39_24_25_46_47_40_41_10_11;TICK;CX_31_39_1_5_12_6_26_18_24_16_14_8_29_37_11_19_30_21_40_34_3_7_10_4_42_36_15_22_2_0_43_47_9_17_13_20_27_35;TICK;M_37_17_47_44_22_5_39_19_32_35_7_20;MX_24_14_33_10_30_26_40_45_2_12_42;DT(6,5,7)rec[-23]_rec[-36];DT(5,2,7)rec[-22];DT(8,5,7)rec[-21]_rec[-35]_rec[-47];DT(8,7,7)rec[-20]_rec[-42];DT(4,7,7)rec[-19]_rec[-46];DT(3,4,7)rec[-18];DT(7,8,7)rec[-17]_rec[-41]_rec[-44];DT(5,4,7)rec[-16]_rec[-40]_rec[-45];DT(6,9,7)rec[-15]_rec[-39];DT(6,3,7)rec[-14]_rec[-43];DT(2,5,7)rec[-13]_rec[-38];DT(4,5,7)rec[-12]_rec[-36]_rec[-37]_rec[-40]_rec[-44]_rec[-64]_rec[-85]_rec[-110];DT(6,1,7)rec[-11]_rec[-32]_rec[-34];DT(4,7,8)rec[-10];DT(7,2,7)rec[-9]_rec[-31];DT(4,3,7)rec[-8]_rec[-28];DT(5,8,7)rec[-7]_rec[-30];DT(7,4,7)rec[-5]_rec[-27]_rec[-29];DT(9,4,7)rec[-4]_rec[-26];DT(2,5,8)rec[-3];DT(3,6,7)rec[-2]_rec[-25];DT(7,6,7)rec[-1]_rec[-33];TICK;R_37_17_47_44_22_5_39_19_32_35_7_20;RX_24_14_33_10_30_26_40_45_2_12_42;TICK;CX_31_39_1_5_12_6_26_18_24_16_14_8_29_37_11_19_30_21_40_34_3_7_10_4_42_36_15_22_2_0_43_47_9_17_13_20_27_35;TICK;CX_2_3_6_7_18_19_14_15_28_29_21_22_12_13_30_31_26_27_34_35_4_5_36_37_42_43_16_17_38_39_24_25_46_47_40_41_10_11;TICK;CX_22_23_25_26_41_42_33_34_20_21_45_46_29_30_31_32_1_2_9_10_5_6_27_28_37_38_17_18_7_8_11_12_35_36_43_44_13_14;TICK;CX_33_25_6_12_18_26_38_44_45_41_8_14_19_11_37_29_22_15_7_3_23_32_4_10_17_9_0_2_20_13_21_30_35_27_5_1_36_42;TICK;M_47_14_10_30_26_44_39_32_2_12_42;MX_24_37_17_33_22_40_5_19_35_45_7_20;DT(9,6,9)rec[-23]_rec[-44];DT(3,6,9)rec[-22]_rec[-36];DT(4,3,9)rec[-21];DT(5,6,9)rec[-20]_rec[-35];DT(5,2,9)rec[-19]_rec[-45];DT(8,7,9)rec[-18]_rec[-43]_rec[-46];DT(7,8,9)rec[-17]_rec[-40];DT(6,9,9)rec[-16]_rec[-38]_rec[-42];DT(2,5,9)rec[-15];DT(3,4,9)rec[-14]_rec[-41];DT(7,4,9)rec[-13]_rec[-37];DT(6,1,9)rec[-12]_rec[-34];DT(6,7,9)rec[-11]_rec[-30];DT(4.5,2.5,9)rec[-10]_rec[-31];DT(6,3,9)rec[-9]_rec[-29]_rec[-32];DT(5,8,9)rec[-8];DT(8,3,9)rec[-7]_rec[-28];DT(2.5,4.5,9)rec[-6]_rec[-26];DT(7,4,10)rec[-4];DT(8,5,9)rec[-3]_rec[-24]_rec[-27];DT(3,6,10)rec[-2];DT(4,7,9)rec[-1]_rec[-33];TICK;R_47_14_10_30_26_44_39_32_2_12_42;RX_24_37_17_33_22_40_5_19_35_45_7_20;TICK;CX_33_25_6_12_18_26_38_44_45_41_8_14_19_11_37_29_22_15_7_3_23_32_4_10_17_9_0_2_20_13_21_30_35_27_5_1_36_42;TICK;CX_22_23_25_26_41_42_33_34_20_21_45_46_29_30_31_32_1_2_9_10_5_6_27_28_37_38_17_18_7_8_11_12_35_36_43_44_13_14;TICK;CX_2_3_6_7_18_19_14_15_28_29_21_22_12_13_30_31_26_27_34_35_4_5_36_37_42_43_16_17_38_39_24_25_46_47_40_41_10_11;TICK;CX_2_0_40_34_31_39_3_7_27_35_1_5_29_37_43_47_9_17_12_6_24_16_10_4_26_18_42_36_15_22_11_19_30_21_14_8;TICK;M_37_17_47_44_22_5_39_19_32_35_7;MX_24_14_33_10_30_26_40_45_2_12_28_42_20;DT(7,6,11)rec[-24];DT(5,2,11)rec[-23];DT(8,5,11)rec[-22]_rec[-37]_rec[-47];DT(8,7,11)rec[-21]_rec[-42];DT(4,7,11)rec[-20]_rec[-46];DT(3,4,11)rec[-19];DT(7,8,11)rec[-18]_rec[-41]_rec[-44];DT(4,3,11)rec[-17]_rec[-45];DT(6,9,11)rec[-16]_rec[-40];DT(6,3,11)rec[-15]_rec[-43];DT(2,5,11)rec[-14]_rec[-39];DT(6,1,11)rec[-13]_rec[-34]_rec[-36];DT(4,7,12)rec[-12];DT(7,2,11)rec[-11]_rec[-33];DT(3.5,3.5,11)rec[-10]_rec[-30];DT(5,8,11)rec[-9]_rec[-32];DT(5,4,11)rec[-8]_rec[-29];DT(7,4,11)rec[-7]_rec[-28]_rec[-31];DT(9,4,11)rec[-6]_rec[-27];DT(2,5,12)rec[-5];DT(3,6,11)rec[-4]_rec[-26];DT(7.5,5.5,11)rec[-2]_rec[-35];DT(4,5,11)rec[-1]_rec[-3]_rec[-25]_rec[-29]_rec[-49]_rec[-53]_rec[-71]_rec[-98];TICK;R_37_17_47_44_22_5_39_19_32_35_7;RX_24_14_33_10_30_26_40_45_2_12_28_42_20;TICK;CX_2_0_40_34_31_39_3_7_27_35_1_5_29_37_43_47_9_17_12_6_24_16_10_4_26_18_42_36_15_22_11_19_30_21_14_8;TICK;CX_2_3_6_7_18_19_14_15_28_29_21_22_12_13_30_31_26_27_34_35_4_5_36_37_42_43_16_17_38_39_24_25_46_47_40_41_10_11;TICK;CX_22_23_25_26_41_42_33_34_20_21_45_46_29_30_31_32_1_2_9_10_5_6_27_28_37_38_17_18_7_8_11_12_35_36_43_44_13_14;TICK;CX_20_13_21_30_35_27_33_25_37_29_6_12_22_15_4_10_18_26_7_3_5_1_38_44_45_41_23_32_17_9_8_14_36_42_0_2;TICK;M_47_14_10_30_26_44_39_19_32_2_12_28_42;MX_24_37_17_33_22_40_5_35_45_7_20;DT(9,6,13)rec[-24]_rec[-46];DT(3,6,13)rec[-23]_rec[-38];DT(4,3,13)rec[-22];DT(5,2,13)rec[-20]_rec[-47];DT(8,7,13)rec[-19]_rec[-45]_rec[-48];DT(7,8,13)rec[-18]_rec[-42];DT(6,9,13)rec[-16]_rec[-40]_rec[-44];DT(2,5,13)rec[-15];DT(3,4,13)rec[-14]_rec[-43];DT(7,4,13)rec[-12]_rec[-39];DT(6,1,13)rec[-11]_rec[-37];DT(6,7,13)rec[-10]_rec[-33];DT(4.5,2.5,13)rec[-9]_rec[-34];DT(6,3,13)rec[-8]_rec[-32]_rec[-35];DT(5,8,13)rec[-7];DT(8,3,13)rec[-6]_rec[-31];DT(2.5,4.5,13)rec[-5]_rec[-29];DT(6,5,13)rec[-4]_rec[-27];DT(8,5,13)rec[-3]_rec[-26]_rec[-30];DT(3,6,14)rec[-2];DT(5,6,13)rec[-1]_rec[-25]_rec[-36];TICK;R_47_14_10_30_26_44_39_19_32_2_12_28_42;RX_24_37_17_33_22_40_5_35_45_7_20;TICK;CX_20_13_21_30_35_27_33_25_37_29_6_12_22_15_4_10_18_26_7_3_5_1_38_44_45_41_23_32_17_9_8_14_36_42_0_2;TICK;CX_22_23_25_26_41_42_33_34_20_21_45_46_29_30_31_32_1_2_9_10_5_6_27_28_37_38_17_18_7_8_11_12_35_36_43_44_13_14;TICK;CX_2_3_6_7_18_19_14_15_28_29_21_22_12_13_30_31_26_27_34_35_4_5_36_37_42_43_16_17_38_39_24_25_46_47_40_41_10_11;TICK;CX_31_39_1_5_12_6_26_18_24_16_14_8_29_37_11_19_30_21_40_34_3_7_10_4_42_36_15_22_2_0_43_47_9_17_13_20_27_35;TICK;M_37_17_47_44_22_5_39_19_32_35_7_20;MX_24_14_33_10_30_26_40_45_2_12_42;DT(6,5,15)rec[-23]_rec[-36];DT(5,2,15)rec[-22];DT(8,5,15)rec[-21]_rec[-35]_rec[-47];DT(8,7,15)rec[-20]_rec[-42];DT(4,7,15)rec[-19]_rec[-46];DT(3,4,15)rec[-18];DT(7,8,15)rec[-17]_rec[-41]_rec[-44];DT(5,4,15)rec[-16]_rec[-40]_rec[-45];DT(6,9,15)rec[-15]_rec[-39];DT(6,3,15)rec[-14]_rec[-43];DT(2,5,15)rec[-13]_rec[-38];DT(4,5,15)rec[-12]_rec[-36]_rec[-37]_rec[-40]_rec[-44]_rec[-64]_rec[-85]_rec[-110];DT(6,1,15)rec[-11]_rec[-32]_rec[-34];DT(4,7,16)rec[-10];DT(7,2,15)rec[-9]_rec[-31];DT(4,3,15)rec[-8]_rec[-28];DT(5,8,15)rec[-7]_rec[-30];DT(7,4,15)rec[-5]_rec[-27]_rec[-29];DT(9,4,15)rec[-4]_rec[-26];DT(2,5,16)rec[-3];DT(3,6,15)rec[-2]_rec[-25];DT(7,6,15)rec[-1]_rec[-33];TICK;R_37_17_47_44_22_5_39_19_32_35_7_20;RX_24_14_33_10_30_26_40_45_2_12_42;TICK;CX_31_39_1_5_12_6_26_18_24_16_14_8_29_37_11_19_30_21_40_34_3_7_10_4_42_36_15_22_2_0_43_47_9_17_13_20_27_35;TICK;CX_2_3_6_7_18_19_14_15_28_29_21_22_12_13_30_31_26_27_34_35_4_5_36_37_42_43_16_17_38_39_24_25_46_47_40_41_10_11;TICK;CX_22_23_25_26_41_42_33_34_20_21_45_46_29_30_31_32_1_2_9_10_5_6_27_28_37_38_17_18_7_8_11_12_35_36_43_44_13_14;TICK;CX_33_25_6_12_18_26_38_44_45_41_8_14_19_11_37_29_22_15_7_3_23_32_4_10_17_9_0_2_20_13_21_30_35_27_5_1_36_42;TICK;M_47_14_10_30_26_44_39_32_2_12_42;MX_24_37_17_33_22_40_5_19_35_45_7_20_16_9_25_4_18_34_1_11_27_41_0_6_28_36_46_3_13_29_43_8_21_38_15_31_23;DT(9,6,17)rec[-48]_rec[-69];DT(3,6,17)rec[-47]_rec[-61];DT(4,3,17)rec[-46];DT(5,6,17)rec[-45]_rec[-60];DT(5,2,17)rec[-44]_rec[-70];DT(8,7,17)rec[-43]_rec[-68]_rec[-71];DT(7,8,17)rec[-42]_rec[-65];DT(6,9,17)rec[-41]_rec[-63]_rec[-67];DT(2,5,17)rec[-40];DT(3,4,17)rec[-39]_rec[-66];DT(7,4,17)rec[-38]_rec[-62];DT(6,1,17)rec[-37]_rec[-59];DT(6,7,17)rec[-36]_rec[-55];DT(4.5,2.5,17)rec[-35]_rec[-56];DT(6,3,17)rec[-34]_rec[-54]_rec[-57];DT(5,8,17)rec[-33];DT(8,3,17)rec[-32]_rec[-53];DT(2.5,4.5,17)rec[-31]_rec[-51];DT(7,4,18)rec[-29];DT(8,5,17)rec[-28]_rec[-49]_rec[-52];DT(3,6,18)rec[-27];DT(4,7,17)rec[-26]_rec[-58];DT(5,3,17)rec[-21]_rec[-23]_rec[-24]_rec[-25]_rec[-35]_rec[-37];DT(7,3,17)rec[-20]_rec[-23]_rec[-34];DT(3,5,17)rec[-14]_rec[-18]_rec[-19]_rec[-22]_rec[-31];DT(6,5,17)rec[-13]_rec[-17]_rec[-18]_rec[-21]_rec[-30];DT(7,5,17)rec[-12]_rec[-16]_rec[-17]_rec[-20]_rec[-29]_rec[-32];DT(9,5,17)rec[-11]_rec[-16]_rec[-28];DT(2,6,17)rec[-10]_rec[-15];DT(3,7,17)rec[-6]_rec[-9]_rec[-10]_rec[-14]_rec[-27];DT(4,5,17)rec[-5]_rec[-8]_rec[-9]_rec[-13]_rec[-26]_rec[-30]_rec[-50]_rec[-54]_rec[-72]_rec[-99];DT(7,7,17)rec[-4]_rec[-7]_rec[-8]_rec[-12]_rec[-36];DT(4,8,17)rec[-3]_rec[-6];DT(5,9,17)rec[-1]_rec[-2]_rec[-3]_rec[-5]_rec[-33];OI(0)rec[-15]_rec[-19]_rec[-22]_rec[-24]_rec[-25]}

\newcommand{\casebbrokencoupler}{https://algassert.com/crumble\#circuit=Q(1,5)0;Q(2,4)1;Q(2,5)2;Q(2,6)3;Q(3,3)4;Q(3,4)5;Q(3,5)6;Q(3,6)7;Q(3,7)8;Q(4,2)9;Q(4,3)10;Q(4,4)11;Q(4,5)12;Q(4,6)13;Q(4,7)14;Q(4,8)15;Q(5,1)16;Q(5,2)17;Q(5,3)18;Q(5,4)19;Q(5,5)20;Q(5,6)21;Q(5,7)22;Q(5,8)23;Q(5,9)24;Q(6,1)25;Q(6,2)26;Q(6,3)27;Q(6,4)28;Q(6,5)29;Q(6,6)30;Q(6,7)31;Q(6,8)32;Q(6,9)33;Q(7,2)34;Q(7,3)35;Q(7,4)36;Q(7,5)37;Q(7,6)38;Q(7,7)39;Q(7,8)40;Q(8,3)41;Q(8,4)42;Q(8,5)43;Q(8,6)44;Q(8,7)45;Q(9,4)46;Q(9,5)47;Q(9,6)48;R_48_14_10_31_27_45_40_33_2_12_20_29_43;RX_16_9_26_4_18_35_1_11_28_42_0_6_37_47_3_13_30_44_8_22_39_15_32_24_21_25_38_17_34_41_23_5_19_36_46_7;TICK;CX_21_13_37_43_34_26_6_12_18_27_46_42_19_11_39_45_38_30_8_14_7_3_4_10_24_33_17_9_23_15_0_2_36_28_22_31_5_1;TICK;CX_19_20_26_27_44_45_34_35_13_14_21_22_28_29_23_24_1_2_9_10_36_37_32_33_5_6_30_31_42_43_17_18_11_12_38_39_46_47_7_8;TICK;CX_39_40_12_13_18_19_47_48_20_21_14_15_22_23_27_28_35_36_4_5_29_30_37_38_2_3_6_7_16_17_31_32_25_26_43_44_41_42_10_11;TICK;CX_44_48_1_5_13_21_12_6_27_18_25_16_32_40_14_8_11_19_41_35_10_4_30_38_15_23_3_7_2_0_31_22_9_17_43_37_28_36;TICK;M_21_38_17_48_45_23_5_40_19_33_36_7;MX_25_34_14_10_31_27_41_46_2_12_20_29_43;DT(6,1,0)rec[-13];DT(7,2,0)rec[-12];DT(4,7,0)rec[-11];DT(4,3,0)rec[-10];DT(6,7,0)rec[-9];DT(8,3,0)rec[-7];DT(9,4,0)rec[-6];DT(2,5,0)rec[-5];DT(4,5,0)rec[-4];DT(5,5,0)rec[-2]_rec[-3]_rec[-8];DT(8,5,0)rec[-1];TICK;R_21_38_17_48_45_23_5_40_19_33_36_7;RX_25_34_14_10_31_27_41_46_2_12_20_29_43;TICK;CX_44_48_1_5_13_21_12_6_27_18_25_16_32_40_14_8_11_19_41_35_10_4_30_38_15_23_3_7_2_0_31_22_9_17_43_37_28_36;TICK;CX_39_40_12_13_18_19_47_48_20_21_14_15_22_23_27_28_35_36_4_5_29_30_37_38_2_3_6_7_16_17_31_32_25_26_43_44_41_42_10_11;TICK;CX_19_20_26_27_44_45_34_35_13_14_21_22_28_29_23_24_1_2_9_10_36_37_32_33_5_6_30_31_42_43_17_18_11_12_38_39_46_47_7_8;TICK;CX_21_13_37_43_34_26_6_12_18_27_46_42_19_11_39_45_38_30_8_14_7_3_4_10_24_33_17_9_23_15_0_2_36_28_22_31_5_1;TICK;M_48_14_10_31_27_45_40_33_2_12_20_29_43;MX_21_25_38_17_34_41_23_5_19_36_46_7;DT(9,6,1)rec[-25]_rec[-47];DT(3,6,1)rec[-24]_rec[-39];DT(4,3,1)rec[-23];DT(5,2,1)rec[-21]_rec[-48];DT(7,6,1)rec[-20]_rec[-46]_rec[-49];DT(7,8,1)rec[-19]_rec[-43];DT(5,8,1)rec[-18]_rec[-41]_rec[-45];DT(2,5,1)rec[-17];DT(3,4,1)rec[-16]_rec[-44];DT(5,4,1)rec[-14]_rec[-15]_rec[-22]_rec[-42]_rec[-50];DT(7,4,1)rec[-13]_rec[-40];DT(4,7,1)rec[-12]_rec[-36];DT(6,1,1)rec[-11]_rec[-38];DT(6,7,1)rec[-10]_rec[-34];DT(4.5,2.5,1)rec[-9]_rec[-35];DT(6,3,1)rec[-8]_rec[-33]_rec[-37];DT(8,3,1)rec[-7]_rec[-32];DT(5,8,2)rec[-6];DT(2.5,4.5,1)rec[-5]_rec[-30];DT(4,5,1)rec[-4]_rec[-28]_rec[-29];DT(6,5,1)rec[-3]_rec[-27];DT(8,5,1)rec[-2]_rec[-26]_rec[-31];DT(3,6,2)rec[-1];TICK;R_48_14_10_31_27_45_40_33_2_12_20_29_43;RX_21_25_38_17_34_41_23_5_19_36_46_7;TICK;CX_21_13_37_43_34_26_6_12_18_27_46_42_19_11_39_45_38_30_8_14_7_3_4_10_24_33_17_9_23_15_0_2_36_28_22_31_5_1;TICK;CX_19_20_26_27_44_45_34_35_13_14_21_22_28_29_23_24_1_2_9_10_36_37_32_33_5_6_30_31_42_43_17_18_11_12_38_39_46_47_7_8;TICK;CX_39_40_12_13_18_19_47_48_20_21_14_15_22_23_27_28_35_36_4_5_29_30_37_38_2_3_6_7_16_17_31_32_25_26_43_44_41_42_10_11;TICK;CX_44_48_1_5_13_21_12_6_27_18_25_16_32_40_14_8_11_19_41_35_10_4_30_38_15_23_3_7_2_0_31_22_9_17_43_37_28_36;TICK;M_21_38_17_48_45_23_5_40_19_33_36_7;MX_25_34_14_10_31_27_41_46_2_12_20_29_43;DT(4,5,3)rec[-25]_rec[-40]_rec[-41];DT(6,5,3)rec[-24]_rec[-39];DT(5,2,3)rec[-23];DT(8,5,3)rec[-22]_rec[-38]_rec[-50];DT(8,7,3)rec[-21]_rec[-45];DT(4,7,3)rec[-20]_rec[-49];DT(3,4,3)rec[-19];DT(6,7,3)rec[-18]_rec[-44]_rec[-47];DT(4,3,3)rec[-17]_rec[-48];DT(6,9,3)rec[-16]_rec[-43];DT(6,3,3)rec[-15]_rec[-46];DT(2,5,3)rec[-14]_rec[-42];DT(6,1,3)rec[-13]_rec[-34]_rec[-36];DT(7,2,3)rec[-12]_rec[-33];DT(4,7,4)rec[-11];DT(3.5,3.5,3)rec[-10]_rec[-30];DT(5,8,3)rec[-9]_rec[-31];DT(7,4,3)rec[-7]_rec[-28]_rec[-32];DT(9,4,3)rec[-6]_rec[-27];DT(2,5,4)rec[-5];DT(3,6,3)rec[-4]_rec[-26];DT(5,4,3)rec[-2]_rec[-3]_rec[-8]_rec[-29]_rec[-37];DT(7,6,3)rec[-1]_rec[-35];TICK;R_21_38_17_48_45_23_5_40_19_33_36_7;RX_25_34_14_10_31_27_41_46_2_12_20_29_43;TICK;CX_44_48_1_5_13_21_12_6_27_18_25_16_32_40_14_8_11_19_41_35_10_4_30_38_15_23_3_7_2_0_31_22_9_17_43_37_28_36;TICK;CX_39_40_12_13_18_19_47_48_20_21_14_15_22_23_27_28_35_36_4_5_29_30_37_38_2_3_6_7_16_17_31_32_25_26_43_44_41_42_10_11;TICK;CX_19_20_26_27_44_45_34_35_13_14_21_22_28_29_23_24_1_2_9_10_36_37_32_33_5_6_30_31_42_43_17_18_11_12_38_39_46_47_7_8;TICK;CX_21_13_37_43_34_26_6_12_18_27_46_42_19_11_39_45_38_30_8_14_7_3_4_10_24_33_17_9_23_15_0_2_36_28_22_31_5_1;TICK;M_48_14_10_31_27_45_40_33_2_12_20_29_43;MX_21_25_38_17_34_41_23_5_19_36_46_7;DT(9,6,5)rec[-25]_rec[-47];DT(3,6,5)rec[-24]_rec[-39];DT(4,3,5)rec[-23];DT(5,2,5)rec[-21]_rec[-48];DT(7,6,5)rec[-20]_rec[-46]_rec[-49];DT(7,8,5)rec[-19]_rec[-43];DT(5,8,5)rec[-18]_rec[-41]_rec[-45];DT(2,5,5)rec[-17];DT(3,4,5)rec[-16]_rec[-44];DT(5,4,5)rec[-14]_rec[-15]_rec[-22]_rec[-42]_rec[-50];DT(7,4,5)rec[-13]_rec[-40];DT(4,7,5)rec[-12]_rec[-36];DT(6,1,5)rec[-11]_rec[-38];DT(6,7,5)rec[-10]_rec[-34];DT(4.5,2.5,5)rec[-9]_rec[-35];DT(6,3,5)rec[-8]_rec[-33]_rec[-37];DT(8,3,5)rec[-7]_rec[-32];DT(5,8,6)rec[-6];DT(2.5,4.5,5)rec[-5]_rec[-30];DT(4,5,5)rec[-4]_rec[-28]_rec[-29];DT(6,5,5)rec[-3]_rec[-27];DT(8,5,5)rec[-2]_rec[-26]_rec[-31];DT(3,6,6)rec[-1];TICK;R_48_14_10_31_27_45_40_33_2_12_20_29_43;RX_21_25_38_17_34_41_23_5_19_36_46_7;TICK;CX_21_13_37_43_34_26_6_12_18_27_46_42_19_11_39_45_38_30_8_14_7_3_4_10_24_33_17_9_23_15_0_2_36_28_22_31_5_1;TICK;CX_19_20_26_27_44_45_34_35_13_14_21_22_28_29_23_24_1_2_9_10_36_37_32_33_5_6_30_31_42_43_17_18_11_12_38_39_46_47_7_8;TICK;CX_39_40_12_13_18_19_47_48_20_21_14_15_22_23_27_28_35_36_4_5_29_30_37_38_2_3_6_7_16_17_31_32_25_26_43_44_41_42_10_11;TICK;CX_44_48_1_5_13_21_12_6_27_18_25_16_32_40_14_8_11_19_41_35_10_4_30_38_15_23_3_7_2_0_31_22_9_17_43_37_28_36;TICK;M_21_38_17_48_45_23_5_40_19_33_36_7;MX_25_34_14_10_31_27_41_46_2_12_20_29_43;DT(4,5,7)rec[-25]_rec[-40]_rec[-41];DT(6,5,7)rec[-24]_rec[-39];DT(5,2,7)rec[-23];DT(8,5,7)rec[-22]_rec[-38]_rec[-50];DT(8,7,7)rec[-21]_rec[-45];DT(4,7,7)rec[-20]_rec[-49];DT(3,4,7)rec[-19];DT(6,7,7)rec[-18]_rec[-44]_rec[-47];DT(4,3,7)rec[-17]_rec[-48];DT(6,9,7)rec[-16]_rec[-43];DT(6,3,7)rec[-15]_rec[-46];DT(2,5,7)rec[-14]_rec[-42];DT(6,1,7)rec[-13]_rec[-34]_rec[-36];DT(7,2,7)rec[-12]_rec[-33];DT(4,7,8)rec[-11];DT(3.5,3.5,7)rec[-10]_rec[-30];DT(5,8,7)rec[-9]_rec[-31];DT(7,4,7)rec[-7]_rec[-28]_rec[-32];DT(9,4,7)rec[-6]_rec[-27];DT(2,5,8)rec[-5];DT(3,6,7)rec[-4]_rec[-26];DT(5,4,7)rec[-2]_rec[-3]_rec[-8]_rec[-29]_rec[-37];DT(7,6,7)rec[-1]_rec[-35];TICK;R_21_38_17_48_45_23_5_40_19_33_36_7;RX_25_34_14_10_31_27_41_46_2_12_20_29_43;TICK;CX_44_48_1_5_13_21_12_6_27_18_25_16_32_40_14_8_11_19_41_35_10_4_30_38_15_23_3_7_2_0_31_22_9_17_43_37_28_36;TICK;CX_39_40_12_13_18_19_47_48_20_21_14_15_22_23_27_28_35_36_4_5_29_30_37_38_2_3_6_7_16_17_31_32_25_26_43_44_41_42_10_11;TICK;CX_19_20_26_27_44_45_34_35_13_14_21_22_28_29_23_24_1_2_9_10_36_37_32_33_5_6_30_31_42_43_17_18_11_12_38_39_46_47_7_8;TICK;CX_21_13_37_43_34_26_6_12_18_27_46_42_19_11_39_45_38_30_8_14_7_3_4_10_24_33_17_9_23_15_0_2_36_28_22_31_5_1;TICK;M_48_14_10_31_27_45_40_33_2_12_20_29_43;MX_21_25_38_17_34_41_23_5_19_36_46_7;DT(9,6,9)rec[-25]_rec[-47];DT(3,6,9)rec[-24]_rec[-39];DT(4,3,9)rec[-23];DT(5,2,9)rec[-21]_rec[-48];DT(7,6,9)rec[-20]_rec[-46]_rec[-49];DT(7,8,9)rec[-19]_rec[-43];DT(5,8,9)rec[-18]_rec[-41]_rec[-45];DT(2,5,9)rec[-17];DT(3,4,9)rec[-16]_rec[-44];DT(5,4,9)rec[-14]_rec[-15]_rec[-22]_rec[-42]_rec[-50];DT(7,4,9)rec[-13]_rec[-40];DT(4,7,9)rec[-12]_rec[-36];DT(6,1,9)rec[-11]_rec[-38];DT(6,7,9)rec[-10]_rec[-34];DT(4.5,2.5,9)rec[-9]_rec[-35];DT(6,3,9)rec[-8]_rec[-33]_rec[-37];DT(8,3,9)rec[-7]_rec[-32];DT(5,8,10)rec[-6];DT(2.5,4.5,9)rec[-5]_rec[-30];DT(4,5,9)rec[-4]_rec[-28]_rec[-29];DT(6,5,9)rec[-3]_rec[-27];DT(8,5,9)rec[-2]_rec[-26]_rec[-31];DT(3,6,10)rec[-1];TICK;R_48_14_10_31_27_45_40_33_2_12_20_29_43;RX_21_25_38_17_34_41_23_5_19_36_46_7;TICK;CX_21_13_37_43_34_26_6_12_18_27_46_42_19_11_39_45_38_30_8_14_7_3_4_10_24_33_17_9_23_15_0_2_36_28_22_31_5_1;TICK;CX_19_20_26_27_44_45_34_35_13_14_21_22_28_29_23_24_1_2_9_10_36_37_32_33_5_6_30_31_42_43_17_18_11_12_38_39_46_47_7_8;TICK;CX_39_40_12_13_18_19_47_48_20_21_14_15_22_23_27_28_35_36_4_5_29_30_37_38_2_3_6_7_16_17_31_32_25_26_43_44_41_42_10_11;TICK;CX_44_48_1_5_13_21_12_6_27_18_25_16_32_40_14_8_11_19_41_35_10_4_30_38_15_23_3_7_2_0_31_22_9_17_43_37_28_36;TICK;M_21_38_17_48_45_23_5_40_19_33_36_7;MX_25_34_14_10_31_27_41_46_2_12_20_29_43;DT(4,5,11)rec[-25]_rec[-40]_rec[-41];DT(6,5,11)rec[-24]_rec[-39];DT(5,2,11)rec[-23];DT(8,5,11)rec[-22]_rec[-38]_rec[-50];DT(8,7,11)rec[-21]_rec[-45];DT(4,7,11)rec[-20]_rec[-49];DT(3,4,11)rec[-19];DT(6,7,11)rec[-18]_rec[-44]_rec[-47];DT(4,3,11)rec[-17]_rec[-48];DT(6,9,11)rec[-16]_rec[-43];DT(6,3,11)rec[-15]_rec[-46];DT(2,5,11)rec[-14]_rec[-42];DT(6,1,11)rec[-13]_rec[-34]_rec[-36];DT(7,2,11)rec[-12]_rec[-33];DT(4,7,12)rec[-11];DT(3.5,3.5,11)rec[-10]_rec[-30];DT(5,8,11)rec[-9]_rec[-31];DT(7,4,11)rec[-7]_rec[-28]_rec[-32];DT(9,4,11)rec[-6]_rec[-27];DT(2,5,12)rec[-5];DT(3,6,11)rec[-4]_rec[-26];DT(5,4,11)rec[-2]_rec[-3]_rec[-8]_rec[-29]_rec[-37];DT(7,6,11)rec[-1]_rec[-35];TICK;R_21_38_17_48_45_23_5_40_19_33_36_7;RX_25_34_14_10_31_27_41_46_2_12_20_29_43;TICK;CX_44_48_1_5_13_21_12_6_27_18_25_16_32_40_14_8_11_19_41_35_10_4_30_38_15_23_3_7_2_0_31_22_9_17_43_37_28_36;TICK;CX_39_40_12_13_18_19_47_48_20_21_14_15_22_23_27_28_35_36_4_5_29_30_37_38_2_3_6_7_16_17_31_32_25_26_43_44_41_42_10_11;TICK;CX_19_20_26_27_44_45_34_35_13_14_21_22_28_29_23_24_1_2_9_10_36_37_32_33_5_6_30_31_42_43_17_18_11_12_38_39_46_47_7_8;TICK;CX_21_13_37_43_34_26_6_12_18_27_46_42_19_11_39_45_38_30_8_14_7_3_4_10_24_33_17_9_23_15_0_2_36_28_22_31_5_1;TICK;M_48_14_10_31_27_45_40_33_2_12_20_29_43;MX_21_25_38_17_34_41_23_5_19_36_46_7;DT(9,6,13)rec[-25]_rec[-47];DT(3,6,13)rec[-24]_rec[-39];DT(4,3,13)rec[-23];DT(5,2,13)rec[-21]_rec[-48];DT(7,6,13)rec[-20]_rec[-46]_rec[-49];DT(7,8,13)rec[-19]_rec[-43];DT(5,8,13)rec[-18]_rec[-41]_rec[-45];DT(2,5,13)rec[-17];DT(3,4,13)rec[-16]_rec[-44];DT(5,4,13)rec[-14]_rec[-15]_rec[-22]_rec[-42]_rec[-50];DT(7,4,13)rec[-13]_rec[-40];DT(4,7,13)rec[-12]_rec[-36];DT(6,1,13)rec[-11]_rec[-38];DT(6,7,13)rec[-10]_rec[-34];DT(4.5,2.5,13)rec[-9]_rec[-35];DT(6,3,13)rec[-8]_rec[-33]_rec[-37];DT(8,3,13)rec[-7]_rec[-32];DT(5,8,14)rec[-6];DT(2.5,4.5,13)rec[-5]_rec[-30];DT(4,5,13)rec[-4]_rec[-28]_rec[-29];DT(6,5,13)rec[-3]_rec[-27];DT(8,5,13)rec[-2]_rec[-26]_rec[-31];DT(3,6,14)rec[-1];TICK;R_48_14_10_31_27_45_40_33_2_12_20_29_43;RX_21_25_38_17_34_41_23_5_19_36_46_7;TICK;CX_21_13_37_43_34_26_6_12_18_27_46_42_19_11_39_45_38_30_8_14_7_3_4_10_24_33_17_9_23_15_0_2_36_28_22_31_5_1;TICK;CX_19_20_26_27_44_45_34_35_13_14_21_22_28_29_23_24_1_2_9_10_36_37_32_33_5_6_30_31_42_43_17_18_11_12_38_39_46_47_7_8;TICK;CX_39_40_12_13_18_19_47_48_20_21_14_15_22_23_27_28_35_36_4_5_29_30_37_38_2_3_6_7_16_17_31_32_25_26_43_44_41_42_10_11;TICK;CX_44_48_1_5_13_21_12_6_27_18_25_16_32_40_14_8_11_19_41_35_10_4_30_38_15_23_3_7_2_0_31_22_9_17_43_37_28_36;TICK;M_21_38_17_48_45_23_5_40_19_33_36_7;MX_25_34_14_10_31_27_41_46_2_12_20_29_43;DT(4,5,15)rec[-25]_rec[-40]_rec[-41];DT(6,5,15)rec[-24]_rec[-39];DT(5,2,15)rec[-23];DT(8,5,15)rec[-22]_rec[-38]_rec[-50];DT(8,7,15)rec[-21]_rec[-45];DT(4,7,15)rec[-20]_rec[-49];DT(3,4,15)rec[-19];DT(6,7,15)rec[-18]_rec[-44]_rec[-47];DT(4,3,15)rec[-17]_rec[-48];DT(6,9,15)rec[-16]_rec[-43];DT(6,3,15)rec[-15]_rec[-46];DT(2,5,15)rec[-14]_rec[-42];DT(6,1,15)rec[-13]_rec[-34]_rec[-36];DT(7,2,15)rec[-12]_rec[-33];DT(4,7,16)rec[-11];DT(3.5,3.5,15)rec[-10]_rec[-30];DT(5,8,15)rec[-9]_rec[-31];DT(7,4,15)rec[-7]_rec[-28]_rec[-32];DT(9,4,15)rec[-6]_rec[-27];DT(2,5,16)rec[-5];DT(3,6,15)rec[-4]_rec[-26];DT(5,4,15)rec[-2]_rec[-3]_rec[-8]_rec[-29]_rec[-37];DT(7,6,15)rec[-1]_rec[-35];TICK;R_21_38_17_48_45_23_5_40_19_33_36_7;RX_25_34_14_10_31_27_41_46_2_12_20_29_43;TICK;CX_44_48_1_5_13_21_12_6_27_18_25_16_32_40_14_8_11_19_41_35_10_4_30_38_15_23_3_7_2_0_31_22_9_17_43_37_28_36;TICK;CX_39_40_12_13_18_19_47_48_20_21_14_15_22_23_27_28_35_36_4_5_29_30_37_38_2_3_6_7_16_17_31_32_25_26_43_44_41_42_10_11;TICK;CX_19_20_26_27_44_45_34_35_13_14_21_22_28_29_23_24_1_2_9_10_36_37_32_33_5_6_30_31_42_43_17_18_11_12_38_39_46_47_7_8;TICK;CX_21_13_37_43_34_26_6_12_18_27_46_42_19_11_39_45_38_30_8_14_7_3_4_10_24_33_17_9_23_15_0_2_36_28_22_31_5_1;TICK;M_48_14_10_31_27_45_40_33_2_12_20_29_43;MX_21_25_38_17_34_41_23_5_19_36_46_7_16_9_26_4_18_35_1_11_28_42_0_6_37_47_3_13_30_44_8_22_39_15_32_24;DT(9,6,17)rec[-49]_rec[-71];DT(3,6,17)rec[-48]_rec[-63];DT(4,3,17)rec[-47];DT(5,2,17)rec[-45]_rec[-72];DT(7,6,17)rec[-44]_rec[-70]_rec[-73];DT(7,8,17)rec[-43]_rec[-67];DT(5,8,17)rec[-42]_rec[-65]_rec[-69];DT(2,5,17)rec[-41];DT(3,4,17)rec[-40]_rec[-68];DT(5,4,17)rec[-38]_rec[-39]_rec[-46]_rec[-66]_rec[-74];DT(7,4,17)rec[-37]_rec[-64];DT(4,7,17)rec[-36]_rec[-60];DT(6,1,17)rec[-35]_rec[-62];DT(6,7,17)rec[-34]_rec[-58];DT(4.5,2.5,17)rec[-33]_rec[-59];DT(6,3,17)rec[-32]_rec[-57]_rec[-61];DT(8,3,17)rec[-31]_rec[-56];DT(5,8,18)rec[-30];DT(2.5,4.5,17)rec[-29]_rec[-54];DT(4,5,17)rec[-28]_rec[-52]_rec[-53];DT(6,5,17)rec[-27]_rec[-51];DT(8,5,17)rec[-26]_rec[-50]_rec[-55];DT(3,6,18)rec[-25];DT(4,2,17)rec[-20]_rec[-22]_rec[-23]_rec[-24]_rec[-33]_rec[-35];DT(6,2,17)rec[-19]_rec[-22]_rec[-32];DT(2,4,17)rec[-13]_rec[-17]_rec[-18]_rec[-21]_rec[-29];DT(6,4,17)rec[-12]_rec[-15]_rec[-16]_rec[-19]_rec[-27]_rec[-31];DT(8,4,17)rec[-11]_rec[-15]_rec[-26];DT(1,5,17)rec[-10]_rec[-14];DT(2,6,17)rec[-6]_rec[-9]_rec[-10]_rec[-13]_rec[-25];DT(4,4,17)rec[-5]_rec[-8]_rec[-9]_rec[-16]_rec[-17]_rec[-20]_rec[-28]_rec[-36];DT(6,6,17)rec[-4]_rec[-7]_rec[-8]_rec[-12]_rec[-34];DT(3,7,17)rec[-3]_rec[-6];DT(4,8,17)rec[-1]_rec[-2]_rec[-3]_rec[-5]_rec[-30];OI(0)rec[-14]_rec[-18]_rec[-21]_rec[-23]_rec[-24]}

\newcommand{\casecbrokencoupler}{https://algassert.com/crumble\#circuit=Q(1,5)0;Q(2,4)1;Q(2,5)2;Q(2,6)3;Q(3,3)4;Q(3,4)5;Q(3,5)6;Q(3,6)7;Q(3,7)8;Q(4,2)9;Q(4,3)10;Q(4,4)11;Q(4,5)12;Q(4,6)13;Q(4,7)14;Q(4,8)15;Q(5,1)16;Q(5,2)17;Q(5,3)18;Q(5,4)19;Q(5,5)20;Q(5,6)21;Q(5,7)22;Q(5,8)23;Q(5,9)24;Q(6,1)25;Q(6,2)26;Q(6,3)27;Q(6,4)28;Q(6,5)29;Q(6,6)30;Q(6,7)31;Q(6,8)32;Q(6,9)33;Q(7,2)34;Q(7,3)35;Q(7,4)36;Q(7,5)37;Q(7,6)38;Q(7,7)39;Q(7,8)40;Q(8,3)41;Q(8,4)42;Q(8,5)43;Q(8,6)44;Q(8,7)45;Q(9,4)46;Q(9,5)47;Q(9,6)48;R_48_14_10_31_27_45_40_33_2_12_29_43;RX_16_9_26_4_18_35_1_11_28_42_0_6_20_37_47_3_13_30_44_8_22_39_15_32_24_21_25_38_17_34_41_23_5_19_36_46_7;TICK;CX_21_13_37_43_34_26_6_12_18_27_46_42_19_11_39_45_38_30_8_14_7_3_4_10_20_29_24_33_17_9_23_15_0_2_36_28_22_31_5_1;TICK;CX_19_20_26_27_44_45_34_35_13_14_21_22_28_29_23_24_1_2_9_10_36_37_32_33_5_6_30_31_42_43_17_18_11_12_38_39_46_47_7_8;TICK;CX_39_40_12_13_18_19_47_48_14_15_22_23_27_28_35_36_4_5_29_30_37_38_2_3_6_7_16_17_31_32_25_26_43_44_41_42_10_11;TICK;CX_21_13_44_48_1_5_12_6_27_18_25_16_32_40_14_8_11_19_41_35_10_4_30_38_15_23_3_7_2_0_31_22_9_17_43_37_28_36;TICK;M_38_17_48_45_23_5_40_19_33_36_20_7_13;MX_25_34_14_10_31_27_41_46_2_12_43;DT(6,1,0)rec[-11];DT(7,2,0)rec[-10];DT(4,7,0)rec[-9];DT(4,3,0)rec[-8];DT(6,7,0)rec[-7];DT(6,3,0)rec[-6];DT(8,3,0)rec[-5];DT(9,4,0)rec[-4];DT(2,5,0)rec[-3];DT(4,5,0)rec[-2];DT(8,5,0)rec[-1];TICK;R_38_17_48_45_23_5_40_19_33_36_20_7_13;RX_25_34_14_10_31_27_41_46_2_12_43;TICK;CX_21_13_44_48_1_5_12_6_27_18_25_16_32_40_14_8_11_19_41_35_10_4_30_38_15_23_3_7_2_0_31_22_9_17_43_37_28_36;TICK;CX_39_40_12_13_18_19_47_48_14_15_22_23_27_28_35_36_4_5_29_30_37_38_2_3_6_7_16_17_31_32_25_26_43_44_41_42_10_11;TICK;CX_19_20_26_27_44_45_34_35_13_14_21_22_28_29_23_24_1_2_9_10_36_37_32_33_5_6_30_31_42_43_17_18_11_12_38_39_46_47_7_8;TICK;CX_21_13_37_43_34_26_6_12_18_27_46_42_19_11_39_45_38_30_8_14_7_3_4_10_20_29_24_33_17_9_23_15_0_2_36_28_22_31_5_1;TICK;M_48_14_10_31_27_45_40_33_2_12_29_43;MX_21_25_38_17_34_41_23_5_19_36_46_7;DT(9,6,1)rec[-24]_rec[-46];DT(3,6,1)rec[-23]_rec[-36]_rec[-37];DT(4,3,1)rec[-22];DT(6,7,1)rec[-21];DT(5,2,1)rec[-20]_rec[-47];DT(7,6,1)rec[-19]_rec[-45]_rec[-48];DT(7,8,1)rec[-18]_rec[-42];DT(5,8,1)rec[-17]_rec[-40]_rec[-44];DT(2,5,1)rec[-16];DT(3,4,1)rec[-15]_rec[-43];DT(5,4,1)rec[-14]_rec[-38]_rec[-41];DT(7,4,1)rec[-13]_rec[-39];DT(4,7,1)rec[-12]_rec[-33];DT(6,1,1)rec[-11]_rec[-35];DT(6.5,6.5,1)rec[-10]_rec[-31];DT(4.5,2.5,1)rec[-9]_rec[-32];DT(6,3,1)rec[-8]_rec[-30]_rec[-34];DT(8,3,1)rec[-7]_rec[-29];DT(5,8,2)rec[-6];DT(2.5,4.5,1)rec[-5]_rec[-27];DT(7,4,2)rec[-3];DT(8,5,1)rec[-2]_rec[-25]_rec[-28];DT(3,6,2)rec[-1];TICK;R_48_14_10_31_27_45_40_33_2_12_29_43;RX_21_25_38_17_34_41_23_5_19_36_46_7;TICK;CX_21_13_37_43_34_26_6_12_18_27_46_42_19_11_39_45_38_30_8_14_7_3_4_10_20_29_24_33_17_9_23_15_0_2_36_28_22_31_5_1;TICK;CX_19_20_26_27_44_45_34_35_13_14_21_22_28_29_23_24_1_2_9_10_36_37_32_33_5_6_30_31_42_43_17_18_11_12_38_39_46_47_7_8;TICK;CX_39_40_12_13_18_19_47_48_14_15_22_23_27_28_35_36_4_5_29_30_37_38_2_3_6_7_16_17_31_32_25_26_43_44_41_42_10_11;TICK;CX_44_48_1_5_12_6_27_18_25_16_32_40_14_8_11_19_41_35_10_4_29_20_30_38_15_23_3_7_2_0_31_22_9_17_43_37_28_36;TICK;M_38_17_48_45_23_5_40_19_33_36_7;MX_21_25_34_14_10_31_27_41_46_2_12_29_43;DT(6,5,3)rec[-24]_rec[-38];DT(5,2,3)rec[-23];DT(8,5,3)rec[-22]_rec[-37]_rec[-48];DT(8,7,3)rec[-21]_rec[-43];DT(4,7,3)rec[-20]_rec[-47];DT(3,4,3)rec[-19];DT(6,7,3)rec[-18]_rec[-42]_rec[-45];DT(4,3,3)rec[-17]_rec[-46];DT(6,9,3)rec[-16]_rec[-41];DT(6,3,3)rec[-15]_rec[-44];DT(2,5,3)rec[-14]_rec[-40];DT(6,1,3)rec[-12]_rec[-33]_rec[-35];DT(7,2,3)rec[-11]_rec[-32];DT(4,7,4)rec[-10];DT(3.5,3.5,3)rec[-9]_rec[-29];DT(5,8,3)rec[-8]_rec[-30];DT(5,4,3)rec[-7]_rec[-28];DT(7,4,3)rec[-6]_rec[-27]_rec[-31];DT(9,4,3)rec[-5]_rec[-26];DT(2,5,4)rec[-4];DT(3,6,3)rec[-3]_rec[-25];DT(4,5,3)rec[-2]_rec[-13]_rec[-28]_rec[-36]_rec[-50];DT(7,6,3)rec[-1]_rec[-34];TICK;R_38_17_48_45_23_5_40_19_33_36_7;RX_21_25_34_14_10_31_27_41_46_2_12_29_43;TICK;CX_44_48_1_5_12_6_27_18_25_16_32_40_14_8_11_19_41_35_10_4_29_20_30_38_15_23_3_7_2_0_31_22_9_17_43_37_28_36;TICK;CX_39_40_12_13_18_19_47_48_14_15_22_23_27_28_35_36_4_5_29_30_37_38_2_3_6_7_16_17_31_32_25_26_43_44_41_42_10_11;TICK;CX_19_20_26_27_44_45_34_35_13_14_21_22_28_29_23_24_1_2_9_10_36_37_32_33_5_6_30_31_42_43_17_18_11_12_38_39_46_47_7_8;TICK;CX_21_13_37_43_34_26_6_12_18_27_46_42_19_11_39_45_38_30_8_14_7_3_4_10_20_29_24_33_17_9_23_15_0_2_36_28_22_31_5_1;TICK;M_48_14_10_31_27_45_40_33_2_12_29_43;MX_21_25_38_17_34_41_23_5_19_36_46_7;DT(9,6,5)rec[-24]_rec[-46];DT(3,6,5)rec[-23]_rec[-38];DT(4,3,5)rec[-22];DT(5,2,5)rec[-20]_rec[-47];DT(7,6,5)rec[-19]_rec[-45]_rec[-48];DT(7,8,5)rec[-18]_rec[-42];DT(5,8,5)rec[-17]_rec[-40]_rec[-44];DT(2,5,5)rec[-16];DT(3,4,5)rec[-15]_rec[-43];DT(5,4,5)rec[-14]_rec[-41];DT(7,4,5)rec[-13]_rec[-39];DT(4,7,5)rec[-12]_rec[-34]_rec[-37];DT(6,1,5)rec[-11]_rec[-36];DT(6,7,5)rec[-10]_rec[-32];DT(4.5,2.5,5)rec[-9]_rec[-33];DT(6,3,5)rec[-8]_rec[-31]_rec[-35];DT(8,3,5)rec[-7]_rec[-30];DT(5,8,6)rec[-6];DT(2.5,4.5,5)rec[-5]_rec[-28];DT(4,5,5)rec[-4]_rec[-27];DT(6,5,5)rec[-3]_rec[-26];DT(8,5,5)rec[-2]_rec[-25]_rec[-29];DT(3,6,6)rec[-1];TICK;R_48_14_10_31_27_45_40_33_2_12_29_43;RX_21_25_38_17_34_41_23_5_19_36_46_7;TICK;CX_21_13_37_43_34_26_6_12_18_27_46_42_19_11_39_45_38_30_8_14_7_3_4_10_20_29_24_33_17_9_23_15_0_2_36_28_22_31_5_1;TICK;CX_19_20_26_27_44_45_34_35_13_14_21_22_28_29_23_24_1_2_9_10_36_37_32_33_5_6_30_31_42_43_17_18_11_12_38_39_46_47_7_8;TICK;CX_39_40_12_13_18_19_47_48_14_15_22_23_27_28_35_36_4_5_29_30_37_38_2_3_6_7_16_17_31_32_25_26_43_44_41_42_10_11;TICK;CX_21_13_44_48_1_5_12_6_27_18_25_16_32_40_14_8_11_19_41_35_10_4_30_38_15_23_3_7_2_0_31_22_9_17_43_37_28_36;TICK;M_38_17_48_45_23_5_40_19_33_36_20_7_13;MX_25_34_14_10_31_27_41_46_2_12_43;DT(6,5,7)rec[-24]_rec[-38];DT(5,2,7)rec[-23];DT(8,5,7)rec[-22]_rec[-37]_rec[-48];DT(8,7,7)rec[-21]_rec[-43];DT(4,7,7)rec[-20]_rec[-47];DT(3,4,7)rec[-19];DT(6,7,7)rec[-18]_rec[-42]_rec[-45];DT(4,3,7)rec[-17]_rec[-46];DT(6,9,7)rec[-16]_rec[-41];DT(6,3,7)rec[-15]_rec[-44];DT(2,5,7)rec[-13]_rec[-40];DT(4,5,7)rec[-12]_rec[-14]_rec[-39]_rec[-45]_rec[-87];DT(6,1,7)rec[-11]_rec[-33]_rec[-35];DT(7,2,7)rec[-10]_rec[-32];DT(4,7,8)rec[-9];DT(3.5,3.5,7)rec[-8]_rec[-29];DT(5,8,7)rec[-7]_rec[-30];DT(5,4,7)rec[-6]_rec[-28];DT(7,4,7)rec[-5]_rec[-27]_rec[-31];DT(9,4,7)rec[-4]_rec[-26];DT(2,5,8)rec[-3];DT(3,6,7)rec[-2]_rec[-25];DT(7,6,7)rec[-1]_rec[-34];TICK;R_38_17_48_45_23_5_40_19_33_36_20_7_13;RX_25_34_14_10_31_27_41_46_2_12_43;TICK;CX_21_13_44_48_1_5_12_6_27_18_25_16_32_40_14_8_11_19_41_35_10_4_30_38_15_23_3_7_2_0_31_22_9_17_43_37_28_36;TICK;CX_39_40_12_13_18_19_47_48_14_15_22_23_27_28_35_36_4_5_29_30_37_38_2_3_6_7_16_17_31_32_25_26_43_44_41_42_10_11;TICK;CX_19_20_26_27_44_45_34_35_13_14_21_22_28_29_23_24_1_2_9_10_36_37_32_33_5_6_30_31_42_43_17_18_11_12_38_39_46_47_7_8;TICK;CX_21_13_37_43_34_26_6_12_18_27_46_42_19_11_39_45_38_30_8_14_7_3_4_10_20_29_24_33_17_9_23_15_0_2_36_28_22_31_5_1;TICK;M_48_14_10_31_27_45_40_33_2_12_29_43;MX_21_25_38_17_34_41_23_5_19_36_46_7;DT(9,6,9)rec[-24]_rec[-46];DT(3,6,9)rec[-23]_rec[-36]_rec[-37];DT(4,3,9)rec[-22];DT(6,7,9)rec[-21];DT(5,2,9)rec[-20]_rec[-47];DT(7,6,9)rec[-19]_rec[-45]_rec[-48];DT(7,8,9)rec[-18]_rec[-42];DT(5,8,9)rec[-17]_rec[-40]_rec[-44];DT(2,5,9)rec[-16];DT(3,4,9)rec[-15]_rec[-43];DT(5,4,9)rec[-14]_rec[-38]_rec[-41];DT(7,4,9)rec[-13]_rec[-39];DT(4,7,9)rec[-12]_rec[-33];DT(6,1,9)rec[-11]_rec[-35];DT(6.5,6.5,9)rec[-10]_rec[-31];DT(4.5,2.5,9)rec[-9]_rec[-32];DT(6,3,9)rec[-8]_rec[-30]_rec[-34];DT(8,3,9)rec[-7]_rec[-29];DT(5,8,10)rec[-6];DT(2.5,4.5,9)rec[-5]_rec[-27];DT(7,4,10)rec[-3];DT(8,5,9)rec[-2]_rec[-25]_rec[-28];DT(3,6,10)rec[-1];TICK;R_48_14_10_31_27_45_40_33_2_12_29_43;RX_21_25_38_17_34_41_23_5_19_36_46_7;TICK;CX_21_13_37_43_34_26_6_12_18_27_46_42_19_11_39_45_38_30_8_14_7_3_4_10_20_29_24_33_17_9_23_15_0_2_36_28_22_31_5_1;TICK;CX_19_20_26_27_44_45_34_35_13_14_21_22_28_29_23_24_1_2_9_10_36_37_32_33_5_6_30_31_42_43_17_18_11_12_38_39_46_47_7_8;TICK;CX_39_40_12_13_18_19_47_48_14_15_22_23_27_28_35_36_4_5_29_30_37_38_2_3_6_7_16_17_31_32_25_26_43_44_41_42_10_11;TICK;CX_44_48_1_5_12_6_27_18_25_16_32_40_14_8_11_19_41_35_10_4_29_20_30_38_15_23_3_7_2_0_31_22_9_17_43_37_28_36;TICK;M_38_17_48_45_23_5_40_19_33_36_7;MX_21_25_34_14_10_31_27_41_46_2_12_29_43;DT(6,5,11)rec[-24]_rec[-38];DT(5,2,11)rec[-23];DT(8,5,11)rec[-22]_rec[-37]_rec[-48];DT(8,7,11)rec[-21]_rec[-43];DT(4,7,11)rec[-20]_rec[-47];DT(3,4,11)rec[-19];DT(6,7,11)rec[-18]_rec[-42]_rec[-45];DT(4,3,11)rec[-17]_rec[-46];DT(6,9,11)rec[-16]_rec[-41];DT(6,3,11)rec[-15]_rec[-44];DT(2,5,11)rec[-14]_rec[-40];DT(6,1,11)rec[-12]_rec[-33]_rec[-35];DT(7,2,11)rec[-11]_rec[-32];DT(4,7,12)rec[-10];DT(3.5,3.5,11)rec[-9]_rec[-29];DT(5,8,11)rec[-8]_rec[-30];DT(5,4,11)rec[-7]_rec[-28];DT(7,4,11)rec[-6]_rec[-27]_rec[-31];DT(9,4,11)rec[-5]_rec[-26];DT(2,5,12)rec[-4];DT(3,6,11)rec[-3]_rec[-25];DT(5,6,11)rec[-2]_rec[-13]_rec[-28]_rec[-36]_rec[-50]_rec[-84];DT(7,6,11)rec[-1]_rec[-34];TICK;R_38_17_48_45_23_5_40_19_33_36_7;RX_21_25_34_14_10_31_27_41_46_2_12_29_43;TICK;CX_44_48_1_5_12_6_27_18_25_16_32_40_14_8_11_19_41_35_10_4_29_20_30_38_15_23_3_7_2_0_31_22_9_17_43_37_28_36;TICK;CX_39_40_12_13_18_19_47_48_14_15_22_23_27_28_35_36_4_5_29_30_37_38_2_3_6_7_16_17_31_32_25_26_43_44_41_42_10_11;TICK;CX_19_20_26_27_44_45_34_35_13_14_21_22_28_29_23_24_1_2_9_10_36_37_32_33_5_6_30_31_42_43_17_18_11_12_38_39_46_47_7_8;TICK;CX_21_13_37_43_34_26_6_12_18_27_46_42_19_11_39_45_38_30_8_14_7_3_4_10_20_29_24_33_17_9_23_15_0_2_36_28_22_31_5_1;TICK;M_48_14_10_31_27_45_40_33_2_12_29_43;MX_21_25_38_17_34_41_23_5_19_36_46_7;DT(9,6,13)rec[-24]_rec[-46];DT(3,6,13)rec[-23]_rec[-38];DT(4,3,13)rec[-22];DT(5,2,13)rec[-20]_rec[-47];DT(7,6,13)rec[-19]_rec[-45]_rec[-48];DT(7,8,13)rec[-18]_rec[-42];DT(5,8,13)rec[-17]_rec[-40]_rec[-44];DT(2,5,13)rec[-16];DT(3,4,13)rec[-15]_rec[-43];DT(5,4,13)rec[-14]_rec[-41];DT(7,4,13)rec[-13]_rec[-39];DT(4,7,13)rec[-12]_rec[-34]_rec[-37];DT(6,1,13)rec[-11]_rec[-36];DT(6,7,13)rec[-10]_rec[-32];DT(4.5,2.5,13)rec[-9]_rec[-33];DT(6,3,13)rec[-8]_rec[-31]_rec[-35];DT(8,3,13)rec[-7]_rec[-30];DT(5,8,14)rec[-6];DT(2.5,4.5,13)rec[-5]_rec[-28];DT(4,5,13)rec[-4]_rec[-27];DT(6,5,13)rec[-3]_rec[-26];DT(8,5,13)rec[-2]_rec[-25]_rec[-29];DT(3,6,14)rec[-1];TICK;R_48_14_10_31_27_45_40_33_2_12_29_43;RX_21_25_38_17_34_41_23_5_19_36_46_7;TICK;CX_21_13_37_43_34_26_6_12_18_27_46_42_19_11_39_45_38_30_8_14_7_3_4_10_20_29_24_33_17_9_23_15_0_2_36_28_22_31_5_1;TICK;CX_19_20_26_27_44_45_34_35_13_14_21_22_28_29_23_24_1_2_9_10_36_37_32_33_5_6_30_31_42_43_17_18_11_12_38_39_46_47_7_8;TICK;CX_39_40_12_13_18_19_47_48_14_15_22_23_27_28_35_36_4_5_29_30_37_38_2_3_6_7_16_17_31_32_25_26_43_44_41_42_10_11;TICK;CX_21_13_44_48_1_5_12_6_27_18_25_16_32_40_14_8_11_19_41_35_10_4_30_38_15_23_3_7_2_0_31_22_9_17_43_37_28_36;TICK;M_38_17_48_45_23_5_40_19_33_36_20_7_13;MX_25_34_14_10_31_27_41_46_2_12_43;DT(6,5,15)rec[-24]_rec[-38];DT(5,2,15)rec[-23];DT(8,5,15)rec[-22]_rec[-37]_rec[-48];DT(8,7,15)rec[-21]_rec[-43];DT(4,7,15)rec[-20]_rec[-47];DT(3,4,15)rec[-19];DT(6,7,15)rec[-18]_rec[-42]_rec[-45];DT(4,3,15)rec[-17]_rec[-46];DT(6,9,15)rec[-16]_rec[-41];DT(6,3,15)rec[-15]_rec[-44];DT(2,5,15)rec[-13]_rec[-40];DT(4,5,15)rec[-12]_rec[-14]_rec[-39]_rec[-45]_rec[-87];DT(6,1,15)rec[-11]_rec[-33]_rec[-35];DT(7,2,15)rec[-10]_rec[-32];DT(4,7,16)rec[-9];DT(3.5,3.5,15)rec[-8]_rec[-29];DT(5,8,15)rec[-7]_rec[-30];DT(5,4,15)rec[-6]_rec[-28];DT(7,4,15)rec[-5]_rec[-27]_rec[-31];DT(9,4,15)rec[-4]_rec[-26];DT(2,5,16)rec[-3];DT(3,6,15)rec[-2]_rec[-25];DT(7,6,15)rec[-1]_rec[-34];TICK;R_38_17_48_45_23_5_40_19_33_36_20_7_13;RX_25_34_14_10_31_27_41_46_2_12_43;TICK;CX_21_13_44_48_1_5_12_6_27_18_25_16_32_40_14_8_11_19_41_35_10_4_30_38_15_23_3_7_2_0_31_22_9_17_43_37_28_36;TICK;CX_39_40_12_13_18_19_47_48_14_15_22_23_27_28_35_36_4_5_29_30_37_38_2_3_6_7_16_17_31_32_25_26_43_44_41_42_10_11;TICK;CX_19_20_26_27_44_45_34_35_13_14_21_22_28_29_23_24_1_2_9_10_36_37_32_33_5_6_30_31_42_43_17_18_11_12_38_39_46_47_7_8;TICK;CX_21_13_37_43_34_26_6_12_18_27_46_42_19_11_39_45_38_30_8_14_7_3_4_10_20_29_24_33_17_9_23_15_0_2_36_28_22_31_5_1;TICK;M_48_14_10_31_27_45_40_33_2_12_29_43;MX_21_25_38_17_34_41_23_5_19_36_46_7_16_9_26_4_18_35_1_11_28_42_0_6_20_37_47_3_13_30_44_8_22_39_15_32_24;DT(9,6,17)rec[-49]_rec[-71];DT(3,6,17)rec[-48]_rec[-61]_rec[-62];DT(4,3,17)rec[-47];DT(6,7,17)rec[-46];DT(5,2,17)rec[-45]_rec[-72];DT(7,6,17)rec[-44]_rec[-70]_rec[-73];DT(7,8,17)rec[-43]_rec[-67];DT(5,8,17)rec[-42]_rec[-65]_rec[-69];DT(2,5,17)rec[-41];DT(3,4,17)rec[-40]_rec[-68];DT(5,4,17)rec[-39]_rec[-63]_rec[-66];DT(7,4,17)rec[-38]_rec[-64];DT(4,7,17)rec[-37]_rec[-58];DT(6,1,17)rec[-36]_rec[-60];DT(6.5,6.5,17)rec[-35]_rec[-56];DT(4.5,2.5,17)rec[-34]_rec[-57];DT(6,3,17)rec[-33]_rec[-55]_rec[-59];DT(8,3,17)rec[-32]_rec[-54];DT(5,8,18)rec[-31];DT(2.5,4.5,17)rec[-30]_rec[-52];DT(7,4,18)rec[-28];DT(8,5,17)rec[-27]_rec[-50]_rec[-53];DT(3,6,18)rec[-26];DT(4,2,17)rec[-21]_rec[-23]_rec[-24]_rec[-25]_rec[-34]_rec[-36];DT(6,2,17)rec[-20]_rec[-23]_rec[-33];DT(2,4,17)rec[-14]_rec[-18]_rec[-19]_rec[-22]_rec[-30];DT(4,4,17)rec[-13]_rec[-17]_rec[-18]_rec[-21]_rec[-29];DT(6,4,17)rec[-12]_rec[-16]_rec[-17]_rec[-20]_rec[-28]_rec[-32];DT(8,4,17)rec[-11]_rec[-16]_rec[-27];DT(1,5,17)rec[-10]_rec[-15];DT(2,6,17)rec[-6]_rec[-9]_rec[-10]_rec[-14]_rec[-26];DT(5,6,17)rec[-5]_rec[-8]_rec[-9]_rec[-13]_rec[-29]_rec[-37]_rec[-51]_rec[-85];DT(6,6,17)rec[-4]_rec[-7]_rec[-8]_rec[-12]_rec[-35];DT(3,7,17)rec[-3]_rec[-6];DT(4,8,17)rec[-1]_rec[-2]_rec[-3]_rec[-5]_rec[-31];OI(0)rec[-15]_rec[-19]_rec[-22]_rec[-24]_rec[-25]}

\newcommand{\casedbrokencoupler}{https://algassert.com/crumble\#circuit=Q(1,5)0;Q(2,4)1;Q(2,5)2;Q(2,6)3;Q(3,3)4;Q(3,4)5;Q(3,5)6;Q(3,6)7;Q(3,7)8;Q(4,2)9;Q(4,3)10;Q(4,4)11;Q(4,5)12;Q(4,6)13;Q(4,7)14;Q(4,8)15;Q(5,1)16;Q(5,2)17;Q(5,3)18;Q(5,4)19;Q(5,5)20;Q(5,6)21;Q(5,7)22;Q(5,8)23;Q(5,9)24;Q(6,1)25;Q(6,2)26;Q(6,3)27;Q(6,4)28;Q(6,5)29;Q(6,6)30;Q(6,7)31;Q(6,8)32;Q(6,9)33;Q(7,2)34;Q(7,3)35;Q(7,4)36;Q(7,5)37;Q(7,6)38;Q(7,7)39;Q(7,8)40;Q(8,3)41;Q(8,4)42;Q(8,5)43;Q(8,6)44;Q(8,7)45;Q(9,4)46;Q(9,5)47;Q(9,6)48;R_48_14_10_31_27_45_40_19_33_2_12_29_43;RX_16_9_26_4_18_35_1_11_28_42_0_6_20_37_47_3_13_30_44_8_22_39_15_32_24_21_25_38_17_34_41_23_5_36_46_7;TICK;CX_21_13_37_43_34_26_6_12_18_27_46_42_39_45_38_30_8_14_7_3_4_10_20_29_24_33_17_9_23_15_0_2_36_28_22_31_5_1;TICK;CX_26_27_44_45_34_35_13_14_21_22_28_29_23_24_1_2_9_10_36_37_32_33_5_6_30_31_42_43_17_18_11_12_38_39_46_47_7_8;TICK;CX_39_40_12_13_18_19_47_48_20_21_14_15_22_23_27_28_35_36_4_5_29_30_37_38_2_3_6_7_16_17_31_32_25_26_43_44_41_42_10_11;TICK;CX_44_48_1_5_13_21_12_6_27_18_25_16_32_40_14_8_11_19_41_35_10_4_29_20_30_38_15_23_3_7_2_0_31_22_9_17_43_37_28_36;TICK;M_21_38_17_48_45_23_5_40_19_33_36_7;MX_25_34_14_10_31_27_41_46_2_12_29_43;DT(6,1,0)rec[-12];DT(7,2,0)rec[-11];DT(4,7,0)rec[-10];DT(4,3,0)rec[-9];DT(6,7,0)rec[-8];DT(8,3,0)rec[-6];DT(9,4,0)rec[-5];DT(2,5,0)rec[-4];DT(4,5,0)rec[-3];DT(6,5,0)rec[-2];DT(8,5,0)rec[-1];TICK;R_21_38_17_48_45_23_5_40_19_33_36_7;RX_25_34_14_10_31_27_41_46_2_12_29_43;TICK;CX_44_48_1_5_13_21_12_6_27_18_25_16_32_40_14_8_11_19_41_35_10_4_29_20_30_38_15_23_3_7_2_0_31_22_9_17_43_37_28_36;TICK;CX_39_40_12_13_18_19_47_48_20_21_14_15_22_23_27_28_35_36_4_5_29_30_37_38_2_3_6_7_16_17_31_32_25_26_43_44_41_42_10_11;TICK;CX_26_27_44_45_34_35_13_14_21_22_28_29_23_24_1_2_9_10_36_37_32_33_5_6_30_31_42_43_17_18_11_12_38_39_46_47_7_8;TICK;CX_21_13_37_43_34_26_6_12_18_27_46_42_39_45_11_19_38_30_8_14_7_3_4_10_24_33_17_9_23_15_0_2_36_28_22_31_5_1;TICK;M_48_14_10_31_27_45_40_33_2_12_43;MX_21_25_38_17_34_41_23_5_11_36_46_20_7;DT(9,6,1)rec[-24]_rec[-45];DT(3,6,1)rec[-23]_rec[-37];DT(4,3,1)rec[-22];DT(5,6,1)rec[-21]_rec[-48];DT(5,2,1)rec[-20]_rec[-46];DT(7,6,1)rec[-19]_rec[-44]_rec[-47];DT(7,8,1)rec[-18]_rec[-41];DT(5,8,1)rec[-17]_rec[-39]_rec[-43];DT(2,5,1)rec[-16];DT(3,4,1)rec[-15]_rec[-42];DT(7,4,1)rec[-14]_rec[-38];DT(4,7,1)rec[-13]_rec[-34];DT(6,1,1)rec[-12]_rec[-36];DT(6,7,1)rec[-11]_rec[-32];DT(4.5,2.5,1)rec[-10]_rec[-33];DT(6,3,1)rec[-9]_rec[-31]_rec[-35];DT(8,3,1)rec[-8]_rec[-30];DT(5,8,2)rec[-7];DT(2.5,4.5,1)rec[-6]_rec[-28];DT(6,5,1)rec[-4]_rec[-26];DT(8,5,1)rec[-3]_rec[-25]_rec[-29];DT(4,5,1)rec[-2]_rec[-5]_rec[-27]_rec[-31];DT(3,6,2)rec[-1];TICK;R_48_14_10_31_27_45_40_33_2_12_43;RX_21_25_38_17_34_41_23_5_11_36_46_20_7;TICK;CX_21_13_37_43_34_26_6_12_18_27_46_42_39_45_11_19_38_30_8_14_7_3_4_10_24_33_17_9_23_15_0_2_36_28_22_31_5_1;TICK;CX_26_27_44_45_34_35_13_14_21_22_28_29_23_24_1_2_9_10_36_37_32_33_5_6_30_31_42_43_17_18_11_12_38_39_46_47_7_8;TICK;CX_39_40_12_13_18_19_47_48_20_21_14_15_22_23_27_28_35_36_4_5_29_30_37_38_2_3_6_7_16_17_31_32_25_26_43_44_41_42_10_11;TICK;CX_44_48_1_5_13_21_12_6_27_18_25_16_32_40_14_8_11_19_41_35_10_4_29_20_30_38_15_23_3_7_2_0_31_22_9_17_43_37_28_36;TICK;M_21_38_17_48_45_23_5_40_19_33_36_7;MX_25_34_14_10_31_27_41_46_2_12_29_43;DT(7,6,3)rec[-23];DT(5,2,3)rec[-22];DT(8,5,3)rec[-21]_rec[-38]_rec[-48];DT(8,7,3)rec[-20]_rec[-43];DT(4,7,3)rec[-19]_rec[-47];DT(3,4,3)rec[-18];DT(6,7,3)rec[-17]_rec[-42]_rec[-45];DT(4,3,3)rec[-16]_rec[-46];DT(6,9,3)rec[-15]_rec[-41];DT(6,3,3)rec[-14]_rec[-44];DT(2,5,3)rec[-13]_rec[-40];DT(6,1,3)rec[-12]_rec[-34]_rec[-36];DT(7,2,3)rec[-11]_rec[-33];DT(4,7,4)rec[-10];DT(4,4,3)rec[-9]_rec[-29]_rec[-30];DT(5,8,3)rec[-8]_rec[-31];DT(6,3,4)rec[-7];DT(7,4,3)rec[-6]_rec[-28]_rec[-32];DT(9,4,3)rec[-5]_rec[-27];DT(2,5,4)rec[-4];DT(3,6,3)rec[-3]_rec[-25];DT(5,5,3)rec[-2]_rec[-26]_rec[-37];DT(7.5,5.5,3)rec[-1]_rec[-35];TICK;R_21_38_17_48_45_23_5_40_19_33_36_7;RX_25_34_14_10_31_27_41_46_2_12_29_43;TICK;CX_44_48_1_5_13_21_12_6_27_18_25_16_32_40_14_8_11_19_41_35_10_4_29_20_30_38_15_23_3_7_2_0_31_22_9_17_43_37_28_36;TICK;CX_39_40_12_13_18_19_47_48_20_21_14_15_22_23_27_28_35_36_4_5_29_30_37_38_2_3_6_7_16_17_31_32_25_26_43_44_41_42_10_11;TICK;CX_26_27_44_45_34_35_13_14_21_22_28_29_23_24_1_2_9_10_36_37_32_33_5_6_30_31_42_43_17_18_11_12_38_39_46_47_7_8;TICK;CX_21_13_37_43_34_26_6_12_18_27_46_42_39_45_38_30_8_14_7_3_4_10_20_29_24_33_17_9_23_15_0_2_36_28_22_31_5_1;TICK;M_48_14_10_31_27_45_40_19_33_2_12_29_43;MX_21_25_38_17_34_41_23_5_36_46_7;DT(9,6,5)rec[-24]_rec[-45];DT(3,6,5)rec[-23]_rec[-37];DT(4,3,5)rec[-22];DT(5,6,5)rec[-21]_rec[-48];DT(5,2,5)rec[-20]_rec[-46];DT(7,6,5)rec[-19]_rec[-44]_rec[-47];DT(7,8,5)rec[-18]_rec[-41];DT(5,8,5)rec[-16]_rec[-39]_rec[-43];DT(2,5,5)rec[-15];DT(3,4,5)rec[-14]_rec[-42];DT(5,4,5)rec[-13]_rec[-17]_rec[-40]_rec[-48]_rec[-63]_rec[-88];DT(7,4,5)rec[-12]_rec[-38];DT(4,7,5)rec[-11]_rec[-34];DT(6,1,5)rec[-10]_rec[-36];DT(6,7,5)rec[-9]_rec[-32];DT(4.5,2.5,5)rec[-8]_rec[-33];DT(6,3,5)rec[-7]_rec[-31]_rec[-35];DT(8,3,5)rec[-6]_rec[-30];DT(5,8,6)rec[-5];DT(2.5,4.5,5)rec[-4]_rec[-28];DT(6,5,5)rec[-3]_rec[-26];DT(8,5,5)rec[-2]_rec[-25]_rec[-29];DT(3,6,6)rec[-1];TICK;R_48_14_10_31_27_45_40_19_33_2_12_29_43;RX_21_25_38_17_34_41_23_5_36_46_7;TICK;CX_21_13_37_43_34_26_6_12_18_27_46_42_39_45_38_30_8_14_7_3_4_10_20_29_24_33_17_9_23_15_0_2_36_28_22_31_5_1;TICK;CX_26_27_44_45_34_35_13_14_21_22_28_29_23_24_1_2_9_10_36_37_32_33_5_6_30_31_42_43_17_18_11_12_38_39_46_47_7_8;TICK;CX_39_40_12_13_18_19_47_48_20_21_14_15_22_23_27_28_35_36_4_5_29_30_37_38_2_3_6_7_16_17_31_32_25_26_43_44_41_42_10_11;TICK;CX_44_48_1_5_13_21_12_6_27_18_25_16_32_40_14_8_11_19_41_35_10_4_29_20_30_38_15_23_3_7_2_0_31_22_9_17_43_37_28_36;TICK;M_21_38_17_48_45_23_5_40_19_33_36_7;MX_25_34_14_10_31_27_41_46_2_12_29_43;DT(4,5,7)rec[-24]_rec[-38];DT(6,5,7)rec[-23]_rec[-37];DT(5,2,7)rec[-22];DT(8,5,7)rec[-21]_rec[-36]_rec[-48];DT(8,7,7)rec[-20]_rec[-43];DT(4,7,7)rec[-19]_rec[-47];DT(3,4,7)rec[-18];DT(6,7,7)rec[-17]_rec[-42]_rec[-45];DT(4,3,7)rec[-16]_rec[-41]_rec[-46];DT(6,9,7)rec[-15]_rec[-40];DT(6,3,7)rec[-14]_rec[-44];DT(2,5,7)rec[-13]_rec[-39];DT(6,1,7)rec[-12]_rec[-32]_rec[-34];DT(7,2,7)rec[-11]_rec[-31];DT(4,7,8)rec[-10];DT(3.5,3.5,7)rec[-9]_rec[-28];DT(5,8,7)rec[-8]_rec[-29];DT(7,4,7)rec[-6]_rec[-27]_rec[-30];DT(9,4,7)rec[-5]_rec[-26];DT(2,5,8)rec[-4];DT(3,6,7)rec[-3]_rec[-25];DT(5,6,7)rec[-2]_rec[-35];DT(7,6,7)rec[-1]_rec[-33];TICK;R_21_38_17_48_45_23_5_40_19_33_36_7;RX_25_34_14_10_31_27_41_46_2_12_29_43;TICK;CX_44_48_1_5_13_21_12_6_27_18_25_16_32_40_14_8_11_19_41_35_10_4_29_20_30_38_15_23_3_7_2_0_31_22_9_17_43_37_28_36;TICK;CX_39_40_12_13_18_19_47_48_20_21_14_15_22_23_27_28_35_36_4_5_29_30_37_38_2_3_6_7_16_17_31_32_25_26_43_44_41_42_10_11;TICK;CX_26_27_44_45_34_35_13_14_21_22_28_29_23_24_1_2_9_10_36_37_32_33_5_6_30_31_42_43_17_18_11_12_38_39_46_47_7_8;TICK;CX_21_13_37_43_34_26_6_12_18_27_46_42_39_45_11_19_38_30_8_14_7_3_4_10_24_33_17_9_23_15_0_2_36_28_22_31_5_1;TICK;M_48_14_10_31_27_45_40_33_2_12_43;MX_21_25_38_17_34_41_23_5_11_36_46_20_7;DT(9,6,9)rec[-24]_rec[-45];DT(3,6,9)rec[-23]_rec[-37];DT(4,3,9)rec[-22];DT(5,6,9)rec[-21]_rec[-48];DT(5,2,9)rec[-20]_rec[-46];DT(7,6,9)rec[-19]_rec[-44]_rec[-47];DT(7,8,9)rec[-18]_rec[-41];DT(5,8,9)rec[-17]_rec[-39]_rec[-43];DT(2,5,9)rec[-16];DT(3,4,9)rec[-15]_rec[-42];DT(7,4,9)rec[-14]_rec[-38];DT(4,7,9)rec[-13]_rec[-34];DT(6,1,9)rec[-12]_rec[-36];DT(6,7,9)rec[-11]_rec[-32];DT(4.5,2.5,9)rec[-10]_rec[-33];DT(6,3,9)rec[-9]_rec[-31]_rec[-35];DT(8,3,9)rec[-8]_rec[-30];DT(5,8,10)rec[-7];DT(2.5,4.5,9)rec[-6]_rec[-28];DT(6,5,9)rec[-4]_rec[-26];DT(8,5,9)rec[-3]_rec[-25]_rec[-29];DT(4,5,9)rec[-2]_rec[-5]_rec[-27]_rec[-31]_rec[-75];DT(3,6,10)rec[-1];TICK;R_48_14_10_31_27_45_40_33_2_12_43;RX_21_25_38_17_34_41_23_5_11_36_46_20_7;TICK;CX_21_13_37_43_34_26_6_12_18_27_46_42_39_45_11_19_38_30_8_14_7_3_4_10_24_33_17_9_23_15_0_2_36_28_22_31_5_1;TICK;CX_26_27_44_45_34_35_13_14_21_22_28_29_23_24_1_2_9_10_36_37_32_33_5_6_30_31_42_43_17_18_11_12_38_39_46_47_7_8;TICK;CX_39_40_12_13_18_19_47_48_20_21_14_15_22_23_27_28_35_36_4_5_29_30_37_38_2_3_6_7_16_17_31_32_25_26_43_44_41_42_10_11;TICK;CX_44_48_1_5_13_21_12_6_27_18_25_16_32_40_14_8_11_19_41_35_10_4_29_20_30_38_15_23_3_7_2_0_31_22_9_17_43_37_28_36;TICK;M_21_38_17_48_45_23_5_40_19_33_36_7;MX_25_34_14_10_31_27_41_46_2_12_29_43;DT(7,6,11)rec[-23];DT(5,2,11)rec[-22];DT(8,5,11)rec[-21]_rec[-38]_rec[-48];DT(8,7,11)rec[-20]_rec[-43];DT(4,7,11)rec[-19]_rec[-47];DT(3,4,11)rec[-18];DT(6,7,11)rec[-17]_rec[-42]_rec[-45];DT(4,3,11)rec[-16]_rec[-46];DT(6,9,11)rec[-15]_rec[-41];DT(6,3,11)rec[-14]_rec[-44];DT(2,5,11)rec[-13]_rec[-40];DT(6,1,11)rec[-12]_rec[-34]_rec[-36];DT(7,2,11)rec[-11]_rec[-33];DT(4,7,12)rec[-10];DT(4,4,11)rec[-9]_rec[-29]_rec[-30];DT(5,8,11)rec[-8]_rec[-31];DT(6,3,12)rec[-7];DT(7,4,11)rec[-6]_rec[-28]_rec[-32];DT(9,4,11)rec[-5]_rec[-27];DT(2,5,12)rec[-4];DT(3,6,11)rec[-3]_rec[-25];DT(5,5,11)rec[-2]_rec[-26]_rec[-37];DT(7.5,5.5,11)rec[-1]_rec[-35];TICK;R_21_38_17_48_45_23_5_40_19_33_36_7;RX_25_34_14_10_31_27_41_46_2_12_29_43;TICK;CX_44_48_1_5_13_21_12_6_27_18_25_16_32_40_14_8_11_19_41_35_10_4_29_20_30_38_15_23_3_7_2_0_31_22_9_17_43_37_28_36;TICK;CX_39_40_12_13_18_19_47_48_20_21_14_15_22_23_27_28_35_36_4_5_29_30_37_38_2_3_6_7_16_17_31_32_25_26_43_44_41_42_10_11;TICK;CX_26_27_44_45_34_35_13_14_21_22_28_29_23_24_1_2_9_10_36_37_32_33_5_6_30_31_42_43_17_18_11_12_38_39_46_47_7_8;TICK;CX_21_13_37_43_34_26_6_12_18_27_46_42_39_45_38_30_8_14_7_3_4_10_20_29_24_33_17_9_23_15_0_2_36_28_22_31_5_1;TICK;M_48_14_10_31_27_45_40_19_33_2_12_29_43;MX_21_25_38_17_34_41_23_5_36_46_7;DT(9,6,13)rec[-24]_rec[-45];DT(3,6,13)rec[-23]_rec[-37];DT(4,3,13)rec[-22];DT(5,6,13)rec[-21]_rec[-48];DT(5,2,13)rec[-20]_rec[-46];DT(7,6,13)rec[-19]_rec[-44]_rec[-47];DT(7,8,13)rec[-18]_rec[-41];DT(5,8,13)rec[-16]_rec[-39]_rec[-43];DT(2,5,13)rec[-15];DT(3,4,13)rec[-14]_rec[-42];DT(5,4,13)rec[-13]_rec[-17]_rec[-40]_rec[-48]_rec[-63]_rec[-88];DT(7,4,13)rec[-12]_rec[-38];DT(4,7,13)rec[-11]_rec[-34];DT(6,1,13)rec[-10]_rec[-36];DT(6,7,13)rec[-9]_rec[-32];DT(4.5,2.5,13)rec[-8]_rec[-33];DT(6,3,13)rec[-7]_rec[-31]_rec[-35];DT(8,3,13)rec[-6]_rec[-30];DT(5,8,14)rec[-5];DT(2.5,4.5,13)rec[-4]_rec[-28];DT(6,5,13)rec[-3]_rec[-26];DT(8,5,13)rec[-2]_rec[-25]_rec[-29];DT(3,6,14)rec[-1];TICK;R_48_14_10_31_27_45_40_19_33_2_12_29_43;RX_21_25_38_17_34_41_23_5_36_46_7;TICK;CX_21_13_37_43_34_26_6_12_18_27_46_42_39_45_38_30_8_14_7_3_4_10_20_29_24_33_17_9_23_15_0_2_36_28_22_31_5_1;TICK;CX_26_27_44_45_34_35_13_14_21_22_28_29_23_24_1_2_9_10_36_37_32_33_5_6_30_31_42_43_17_18_11_12_38_39_46_47_7_8;TICK;CX_39_40_12_13_18_19_47_48_20_21_14_15_22_23_27_28_35_36_4_5_29_30_37_38_2_3_6_7_16_17_31_32_25_26_43_44_41_42_10_11;TICK;CX_44_48_1_5_13_21_12_6_27_18_25_16_32_40_14_8_11_19_41_35_10_4_29_20_30_38_15_23_3_7_2_0_31_22_9_17_43_37_28_36;TICK;M_21_38_17_48_45_23_5_40_19_33_36_7;MX_25_34_14_10_31_27_41_46_2_12_29_43;DT(4,5,15)rec[-24]_rec[-38];DT(6,5,15)rec[-23]_rec[-37];DT(5,2,15)rec[-22];DT(8,5,15)rec[-21]_rec[-36]_rec[-48];DT(8,7,15)rec[-20]_rec[-43];DT(4,7,15)rec[-19]_rec[-47];DT(3,4,15)rec[-18];DT(6,7,15)rec[-17]_rec[-42]_rec[-45];DT(4,3,15)rec[-16]_rec[-41]_rec[-46];DT(6,9,15)rec[-15]_rec[-40];DT(6,3,15)rec[-14]_rec[-44];DT(2,5,15)rec[-13]_rec[-39];DT(6,1,15)rec[-12]_rec[-32]_rec[-34];DT(7,2,15)rec[-11]_rec[-31];DT(4,7,16)rec[-10];DT(3.5,3.5,15)rec[-9]_rec[-28];DT(5,8,15)rec[-8]_rec[-29];DT(7,4,15)rec[-6]_rec[-27]_rec[-30];DT(9,4,15)rec[-5]_rec[-26];DT(2,5,16)rec[-4];DT(3,6,15)rec[-3]_rec[-25];DT(5,6,15)rec[-2]_rec[-35];DT(7,6,15)rec[-1]_rec[-33];TICK;R_21_38_17_48_45_23_5_40_19_33_36_7;RX_25_34_14_10_31_27_41_46_2_12_29_43;TICK;CX_44_48_1_5_13_21_12_6_27_18_25_16_32_40_14_8_11_19_41_35_10_4_29_20_30_38_15_23_3_7_2_0_31_22_9_17_43_37_28_36;TICK;CX_39_40_12_13_18_19_47_48_20_21_14_15_22_23_27_28_35_36_4_5_29_30_37_38_2_3_6_7_16_17_31_32_25_26_43_44_41_42_10_11;TICK;CX_26_27_44_45_34_35_13_14_21_22_28_29_23_24_1_2_9_10_36_37_32_33_5_6_30_31_42_43_17_18_11_12_38_39_46_47_7_8;TICK;CX_21_13_37_43_34_26_6_12_18_27_46_42_39_45_11_19_38_30_8_14_7_3_4_10_24_33_17_9_23_15_0_2_36_28_22_31_5_1;TICK;M_48_14_10_31_27_45_40_33_2_12_43;MX_21_25_38_17_34_41_23_5_11_36_46_20_7_16_9_26_4_18_35_1_19_28_42_0_6_29_37_47_3_13_30_44_8_22_39_15_32_24;DT(9,6,17)rec[-49]_rec[-70];DT(3,6,17)rec[-48]_rec[-62];DT(4,3,17)rec[-47];DT(5,6,17)rec[-46]_rec[-73];DT(5,2,17)rec[-45]_rec[-71];DT(7,6,17)rec[-44]_rec[-69]_rec[-72];DT(7,8,17)rec[-43]_rec[-66];DT(5,8,17)rec[-42]_rec[-64]_rec[-68];DT(2,5,17)rec[-41];DT(3,4,17)rec[-40]_rec[-67];DT(7,4,17)rec[-39]_rec[-63];DT(4,7,17)rec[-38]_rec[-59];DT(6,1,17)rec[-37]_rec[-61];DT(6,7,17)rec[-36]_rec[-57];DT(4.5,2.5,17)rec[-35]_rec[-58];DT(6,3,17)rec[-34]_rec[-56]_rec[-60];DT(8,3,17)rec[-33]_rec[-55];DT(5,8,18)rec[-32];DT(2.5,4.5,17)rec[-31]_rec[-53];DT(6,5,17)rec[-29]_rec[-51];DT(8,5,17)rec[-28]_rec[-50]_rec[-54];DT(4,5,17)rec[-27]_rec[-30]_rec[-52]_rec[-56]_rec[-100];DT(3,6,18)rec[-26];DT(4,2,17)rec[-21]_rec[-23]_rec[-24]_rec[-25]_rec[-35]_rec[-37];DT(6,2,17)rec[-20]_rec[-23]_rec[-34];DT(2,4,17)rec[-14]_rec[-18]_rec[-19]_rec[-22]_rec[-30]_rec[-31];DT(5,3,17)rec[-13]_rec[-17]_rec[-18]_rec[-21];DT(6,4,17)rec[-12]_rec[-16]_rec[-17]_rec[-20]_rec[-29]_rec[-33];DT(8,4,17)rec[-11]_rec[-16]_rec[-28];DT(1,5,17)rec[-10]_rec[-15];DT(2,6,17)rec[-6]_rec[-9]_rec[-10]_rec[-14]_rec[-26];DT(4,6,17)rec[-5]_rec[-8]_rec[-9]_rec[-13]_rec[-27]_rec[-38];DT(6,6,17)rec[-4]_rec[-7]_rec[-8]_rec[-12]_rec[-36];DT(3,7,17)rec[-3]_rec[-6];DT(4,8,17)rec[-1]_rec[-2]_rec[-3]_rec[-5]_rec[-32];OI(0)rec[-15]_rec[-19]_rec[-22]_rec[-24]_rec[-25]}

\title{Handling fabrication defects in hex-grid surface codes}

\author{Oscar Higgott}
\email{oscarhiggott@google.com}
\affiliation{%
 Google Quantum AI
}%
\author{Benjamin Anker}%
\affiliation{%
 Google Quantum AI
}%
\affiliation{Center for Quantum Information and Control (CQuIC), University of New Mexico}
\affiliation{Electrical and Computer Engineering Department, University of New Mexico}
\author{Matt McEwen}%
\affiliation{%
 Google Quantum AI
}%
\author{Dripto M. Debroy}%
\affiliation{%
 Google Quantum AI
}%

\date{\today}

\begin{abstract}
Recent work has shown that a hexagonal grid qubit layout, with only three couplers per qubit, is sufficient to implement the surface code with performance comparable to that of a traditional four-coupler layout~\cite{mcewen_relaxing_2023}.
In this work we propose a method for handling broken qubits and couplers even in hex-grid surface code architectures, using an extension of the LUCI framework~\cite{debroy_luci_2024}.
We show that for isolated broken qubits, the circuit distance drops by one, while for isolated broken couplers, the distance drops by one in one or both bases.
By providing a viable dropout strategy, we have removed a critical roadblock to the implementation of hexagonal qubit grids in hardware for large-scale quantum error correction.
\end{abstract}

\maketitle

\section{Introduction}\label{sec:intro}
Quantum error correction (QEC) is essential for enabling scalable quantum computation.
One of the most well-studied QEC codes is the surface code~\cite{dennis2002topological,fowler2012surface}, which has now been implemented below threshold in both superconducting and neutral atom architectures~\cite{Acharya2024qecbelowsurface, bluvstein2025architectural}.
However, scaling up to a large-scale fault-tolerant quantum computer remains an immense challenge, motivating the development of improved QEC codes and circuits that ease hardware requirements.

\begin{figure}[t]
    \centering
    \includegraphics[width=0.95\linewidth]{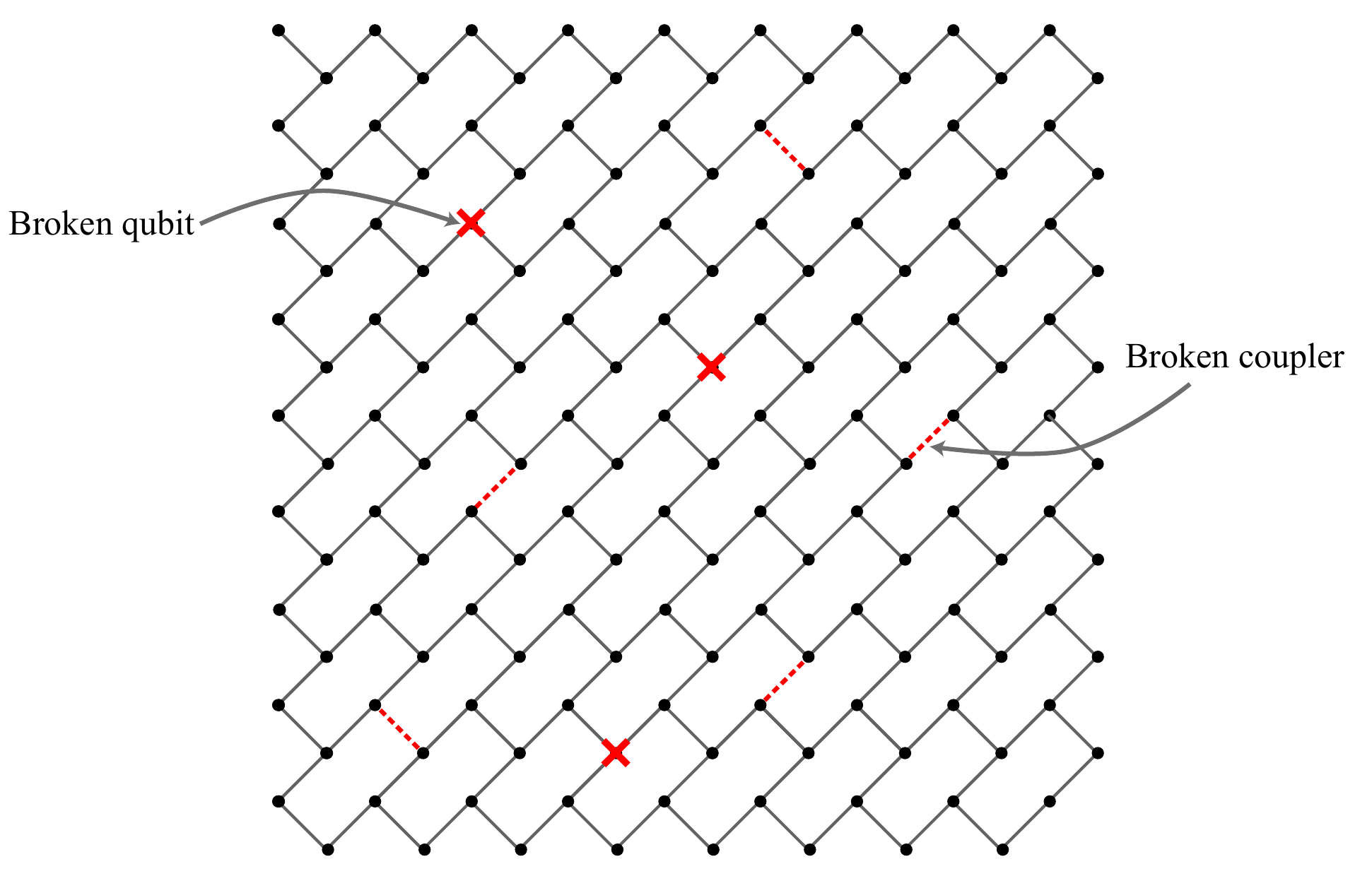}
    \caption{A hex-grid surface code with broken qubits and couplers. Black circles are qubits and black lines are couplers. Broken qubits and couplers are shown in red.
    }
    \label{fig:hex_grid_dropouts}
\end{figure}

Here we tackle two challenges encountered with experimental realizations of the surface code.
Firstly, broken qubits and couplers, as seen in Fig.~\ref{fig:hex_grid_dropouts}, inevitably arise during fabrication in some platforms, requiring modifications to the code or circuit to maintain good performance~\cite{auger2017fault,strikis2023quantum,Siegel2023adaptive, lin2024codesign, wei2024low, debroy_luci_2024, leroux2024snakes, zhou2024halma,aasen2023fault, mclauchlan2024accommodating, GransSamuelsson2024improvedpairwise, mishmash2025excisingdeadcomponentssurface}.
Secondly, the degree-four connectivity of the surface code can be challenging to implement, leading to frequency collisions in some architectures~\cite{chamberland2020topological}, and a larger number of components relative to topologies with lower qubit degree.
This motivates the design of codes and circuits with reduced connectivity requirements~\cite{chamberland2020topological,bravyi2013subsystem, Hastings2021dynamically,bacon2006operator,gidney2023less}.

Recent work has revealed enormous freedom in the underlying implementation of QEC~\cite{mcewen_relaxing_2023}, by considering the spacetime structure of QEC circuits and their detecting regions.
This freedom can be used to implement the surface code on a hex-grid lattice, with only three couplers per qubit, while retaining almost identical performance to standard standard surface code circuits~\cite{mcewen_relaxing_2023, eickbusch2024demonstrating}.
Alternatively, we can use this freedom to adapt circuits to defects, as done in the recently-introduced LUCI framework~\cite{debroy_luci_2024}.
\begin{figure}[b!]
    \centering
    \includegraphics[width=\linewidth]{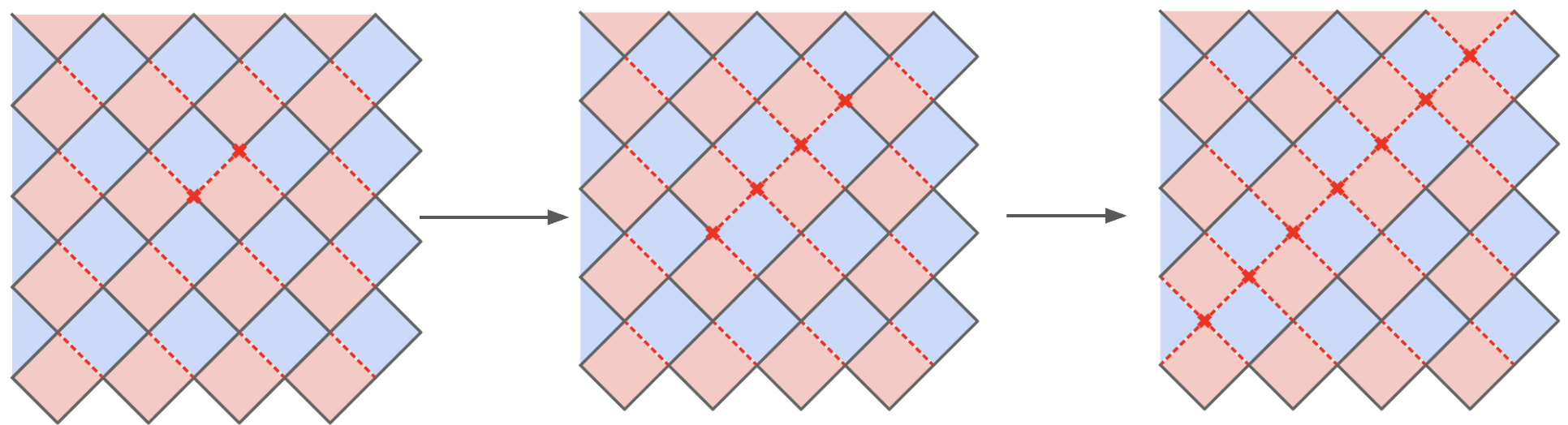}
    \caption{Using the original version of LUCI, a single data qubit error in the hex-grid surface code cascades to a chain of disabled qubits and couplers that spans the lattice.}
    \label{fig:cascading_hex_grid_dropout_in_original_luci}
\end{figure}

However, these two strategies seem at odds: if we implement a hex-grid surface code circuit and experience a dropout, we do not have the same freedom to adapt to broken qubits or couplers.
Indeed, when the original LUCI approach is applied to hex-grid circuits, it is unable to produce a valid circuit if even a single qubit or coupler is broken, as seen in Fig.~\ref{fig:cascading_hex_grid_dropout_in_original_luci}.

\begin{figure*}[t]
    \centering
    \includegraphics[width=\linewidth]{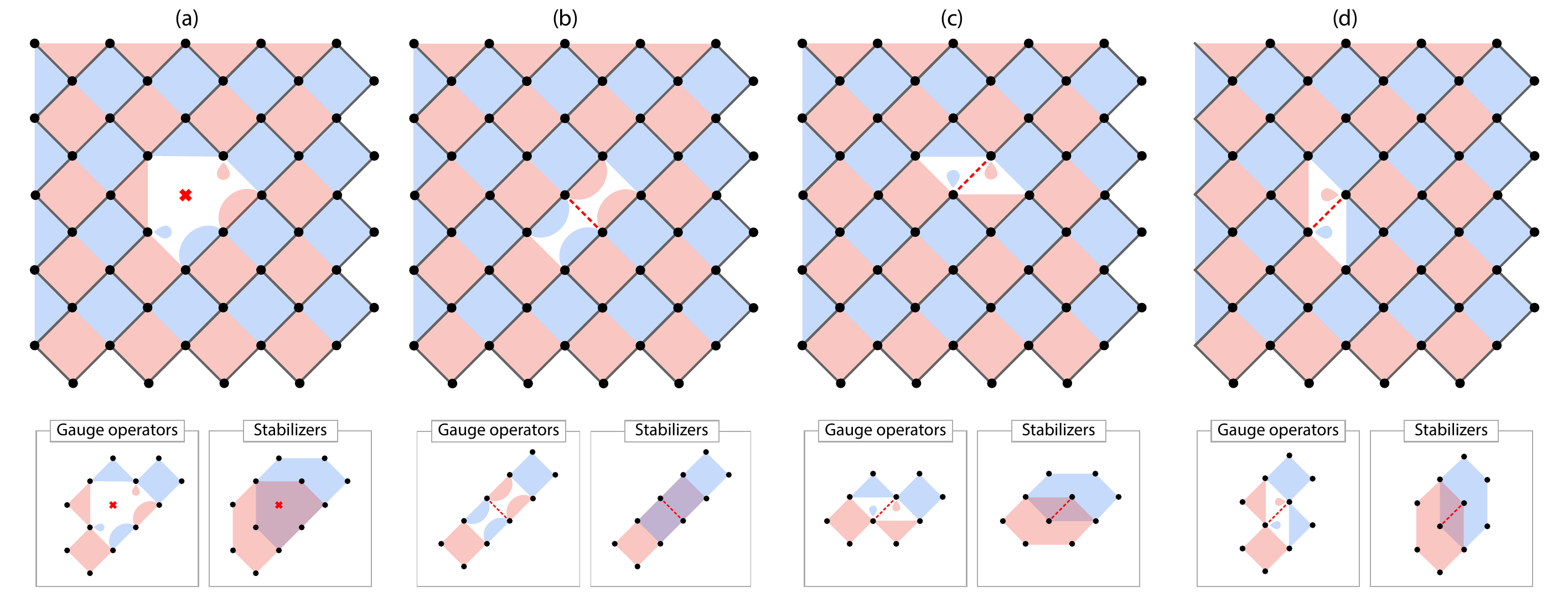}
    \caption{The structure of stabilizers and gauge operators in the mid-cycle subsystem code for our circuits for handling isolated broken qubits or couplers in a hex-grid architecture.
    Links for opening the circuits in Crumble: 
    \href{\caseabrokenqubithex}{Case A, Broken Qubit. }
    \href{\casebbrokencoupler}{Case B, Broken Coupler. }
    \href{\casecbrokencoupler}{Case C, Broken Coupler. }
    \href{\casecbrokencoupler}{Case D, Broken Coupler. }
    }\label{fig:lost_hex_component_examples}
\end{figure*}
Here we present improvements to the LUCI approach for dropouts, which enable us to have the best of both worlds.
Specifically, with our improvements, we show that LUCI can tolerate dropout in hex-grid surface codes.
In this construction, the fault distance only drops by one for any broken qubit. 
For broken couplers, depending on the orientation, the fault distance drops by one in either the $X$ basis, the $Z$ basis or both bases. 

\section{Handling fabrication defects in a hex-grid surface code}\label{sec:dropouts}
Our method for handling dropout in hex-grid surface code circuits uses an extension of the LUCI framework, introduced in \cite{debroy_luci_2024}.
Similarly to other approaches to handling dropout, LUCI modifies the surface code to a subsystem code that does not make use of the missing qubits and couplers.
However, LUCI defines the subsystem code in the \textit{mid-cycle} of the circuit~\cite{mcewen_relaxing_2023}, rather than at the end-cycle.
Concretely, the mid-cycle subsystem code in LUCI is defined on all the qubits (both data and measure qubits).
In each \textit{round} of a LUCI circuit, a subset of the mid-cycle checks (either regular stabilizers or gauge operators) is measured using a constant-depth circuit (a \textit{contracting} circuit) that maps each such check to a single Pauli operator that is then measured.
The measured qubits are then re-initialized and the inverse of the contracting circuit, an \textit{expanding} circuit, is applied to return to the original mid-cycle subsystem code.
We refer the reader to \cite{debroy_luci_2024} for a complete description of LUCI.

One possible approach for handling broken qubits in the hex-grid surface code would be to treat each coupler not present in the hex-grid lattice (relative to the square grid) as a broken coupler, and then apply the original LUCI framework~\cite{debroy_luci_2024}.
Unfortunately, this approach is unable to produce a valid circuit for hex-grid surface code circuits in the presence of broken qubits or couplers.
Specifically, the original LUCI method removes any qubit that has two perpendicular broken couplers attached to it (see step 1 of Figure~7 of \cite{debroy_luci_2024}).
This is because the original LUCI method supports only one gauge operator per plaquette; since a qubit adjacent to two perpendicular broken couplers is isolated from the other qubits in the same plaquette, the qubits in the plaquette cannot all be measured in a single gauge operator measurement.
For the hex-grid surface code circuit, this issue leads to the aforementioned cascading effect.

We address this challenge by changing the structure of the mid-cycle subsystem code (including the introduction of weight-one gauge operators), and allowing multiple gauge operators to be measured within the same plaquette.
We show the structure of the mid-cycle subsystem code with our method for four key cases in \Cref{fig:lost_hex_component_examples}, corresponding to a broken qubit and three different orientations of broken coupler.
These check operators (stabilizers and gauge operators) are measured over the course of four rounds, where each round is a contracting circuit followed by an expanding circuit.

A broken data qubit in the bulk of the lattice results in the configuration of gauge operators in the mid-cycle shown in \Cref{fig:lost_hex_component_examples}(a), consisting of a weight-four, a weight-three, a weight-two and a weight-one gauge operator in each basis ($X$ and $Z$).
The four $X$-type gauge operators are combined to form a weight-8 $X$-type stabilizer in the mid-cycle, and similarly for the $Z$-type gauge operators.
Since these $X$ and $Z$ mid-cycle stabilizers span two plaquettes in both the horizontal and vertical direction, the circuit distance is reduced by one in both the $X$ basis and the $Z$ basis.

\begin{figure*}[t]
    \centering
    \includegraphics[width=\linewidth]{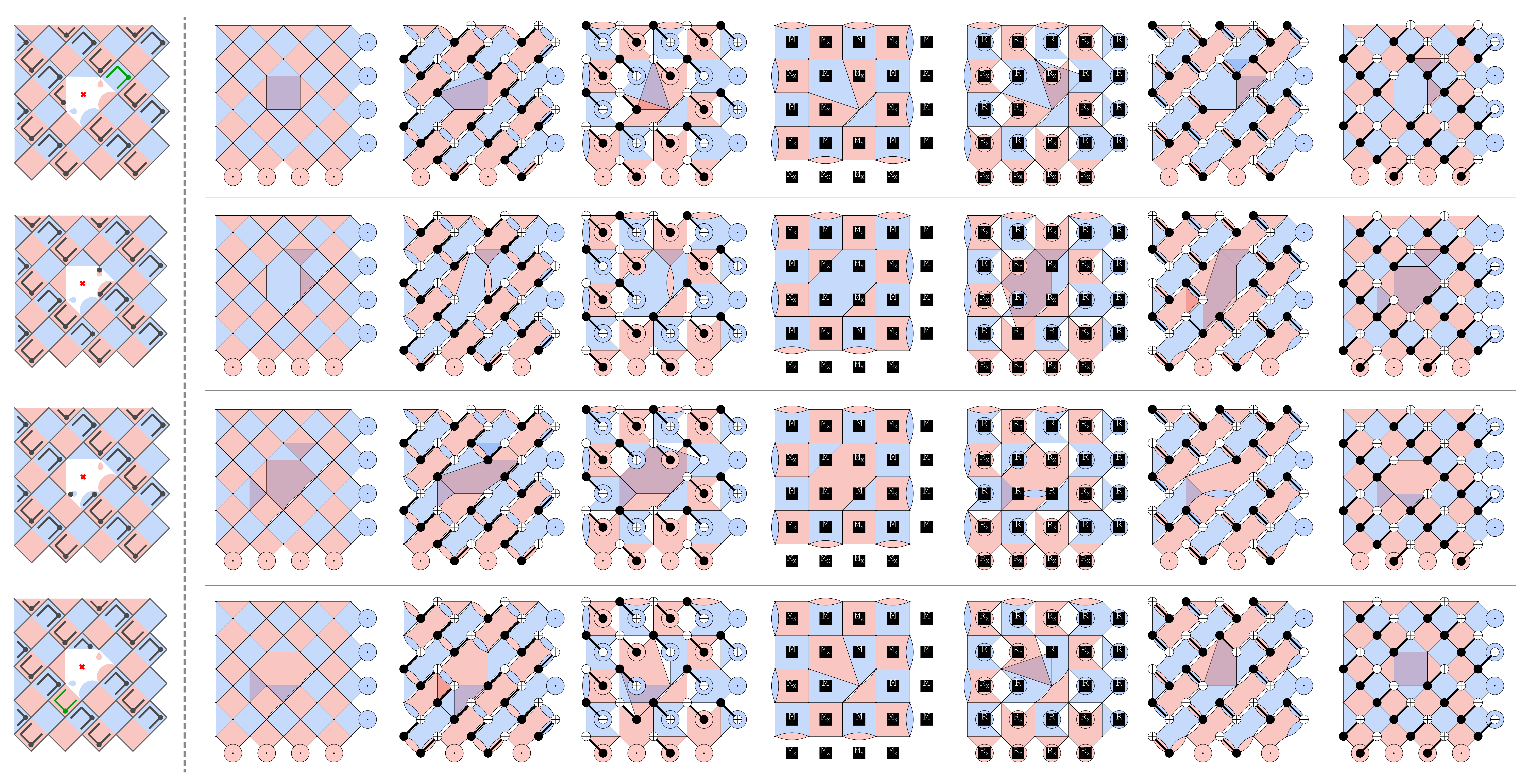}
    \caption{Circuit and detecting regions for handling a broken data qubit in the hex-grid surface code using LUCI. Each row corresponds to one of the four types of rounds in the circuit, with the corresponding LUCI board shown in the far-left column. In the original version of LUCI, each gauge operator is measured once every four rounds. However, we introduce additional gauge operator measurements here (drawn in green) which each introduce an additional small detector (corresponding to a repeated gauge operator measurement).     \href{\caseabrokenqubithex}{Link to the circuit in Crumble.}
    }
    \label{fig:broken_qubit_detecting_regions}
\end{figure*}
We also show how a broken coupler is handled.
In \Cref{fig:lost_hex_component_examples} we show the structure of the gauge operators and stabilizers in the mid-cycle when each of the three couplers adjacent to a qubit in the bulk is broken.
For the example in \Cref{fig:lost_hex_component_examples}(b), both the $X$ distance and the $Z$ distance drop by one, since the gauge operators are combined into a stabilizer which spans two plaquettes in both the horizontal and vertical direction.
For (c) the $Z$ distance drops by one while the $X$ distance is preserved (since the stabilizers are elongated horizontally).
For (d), instead the $X$ distance drops by one and the $Z$ distance is preserved (the stabilizers are elongated vertically).

Independently of the improvements to our choice of mid-cycle subsystem code, we have also made improvements to how the four-round LUCI diagram is constructed (i.e.~the circuit implementation of the mid-cycle subsystem code), as well as the detector inference for the resultant circuit.

Our improvements to the LUCI diagram make use of schedule-induced gauge-fixing~\cite{higgott2021subsystem}, known as \textit{shells}~\cite{strikis2023quantum} in the context of handling fabrication defects.
Specifically, certain gauge operator measurements can be measured while they are still an element of the instantaneous stabilizer group (ISG), resulting in additional small detectors, improving QEC performance.
In the original version of LUCI~\cite{debroy_luci_2024}, each gauge operator is measured exactly once over the course of the four rounds of the LUCI boards.
Concretely, the four rounds of the LUCI diagram have the following structure:
\begin{enumerate}
    \item Measure even diagonals (skip $Z$ gauge operators)
    \item Measure odd diagonals (skip $Z$ gauge operators)
    \item Measure even diagonals (skip $X$ gauge operators)
    \item Measure odd diagonals (skip $X$ gauge operators)
\end{enumerate}
Since this circuit measures all $X$ gauge operators and then all $Z$ gauge operators, it guarantees that each super-stabilizer is measured correctly (since only mutually commuting operators are measured in the two consecutive time steps of the circuit that measure each super-stabilizer).

However, here we show that additional gauge operator measurements can be included which break this structure but improve QEC performance.
Specifically, we start with the usual LUCI board and then in each round we consider each gauge operator $g$ that would usually be skipped due to its Pauli type (i.e.~a $Z$ gauge operator in rounds 1 or 2 or an $X$ gauge operator in rounds 3 or 4) but which is otherwise in the correct diagonal of the mid-cycle subsystem code.
We choose to include $g$ as a measurement in the round provided that it
\begin{enumerate}
    \renewcommand{\theenumi}{\alph{enumi}}
    \item Commutes with all other gauge operators in the same round, and
    \item Commutes with the \textit{product} of gauge operators within each super-stabilizer that were measured in the previous round.
\end{enumerate}
By adding these two constraints, we ensure that adding $g$ does not anti-commute with any of the detectors from the original LUCI diagram (this can be understood using the condition proven in Appendix A of \cite{Suchara_2011_topological_subsystem}).
For the circuits we consider, adding these additional measurements increases the number of determined measurements in the circuit, and therefore the number of detectors, without increasing the overall circuit depth.
For a broken data qubit, these additional gauge operator measurements are shown in green in the left column of \Cref{fig:broken_qubit_detecting_regions}.

Where stabilizers are measured directly the choice of detecting regions is straightforward, in that we can always define a detector as the product of two consecutive stabilizer measurements. 
However, when we consider the gauge operators of a subsystem code, which we must in this setting, the choice is not so clear since the product of two consecutive \emph{gauge operator} measurements is not necessarily deterministic, and we do not always measure an entire stabilizer when we measure one of its constituent gauge operators.

How to choose detecting regions becomes more clear, though, if we instead consider the instantaneous stabilizer group (ISG) \cite{Hastings2021dynamically, gottesman1998heisenberg} relative to the measurements so far performed. In essence, we maintain a basis for the ISG which we update after each round of commuting preparation/measurement. 
The basis we choose is tailored to the choice of gauge operators we made as part of the LUCI construction such that when we measure a gauge operator $g$, we can (efficiently) check to see if it is a member of the chosen basis for the ISG. 
If so, we add the product of the last two times it was measured as a detector. 
Otherwise, we update the basis we have chosen for the ISG such that
\begin{enumerate}
    \item $g$ is one of the basis elements
    \item elements $a_i$ which anticommute with $g$ are paired up into basis elements which commute with $g$
    \item basis elements previously produced by pairing up $g$ with other basis elements $g'g$ are removed and replaced with $g'$.
\end{enumerate}
These improvements mean that the choice of detectors is independent of assumptions about the order of the measurements. In the example grids above, adding the one or two measurements that would have been omitted by the original LUCI construction improved logical error rates by ${\sim}10\%$. 

\section{Results}\label{sec:benchmarking}
\begin{figure}[h]
    \centering
    \includegraphics[width=0.95\linewidth]{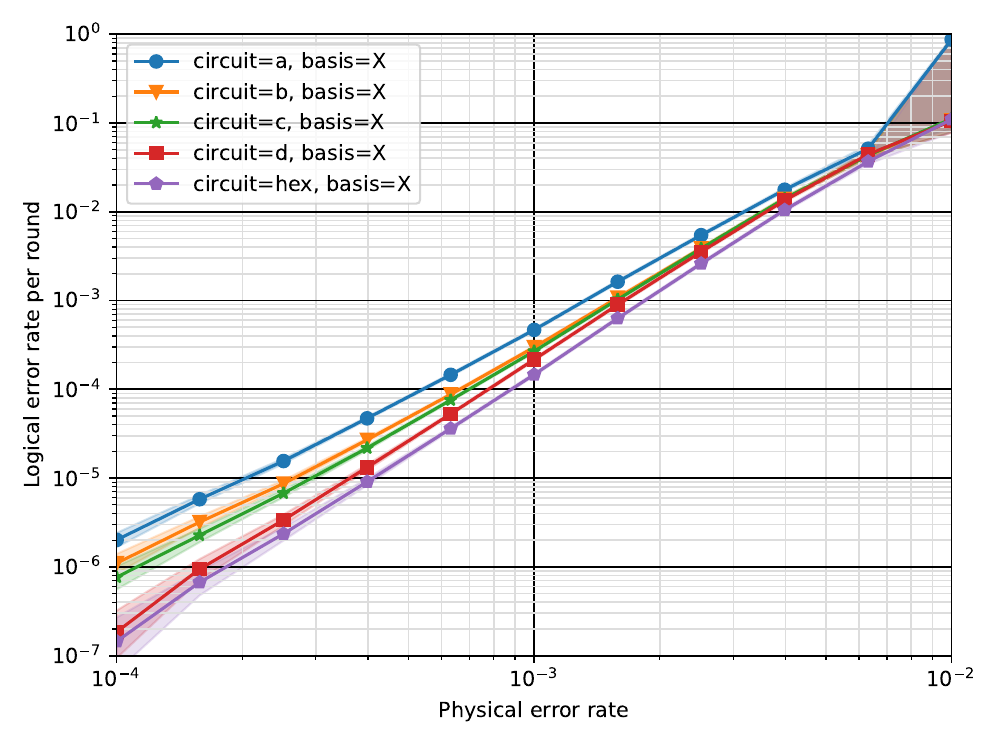}
    \includegraphics[width=0.95\linewidth]{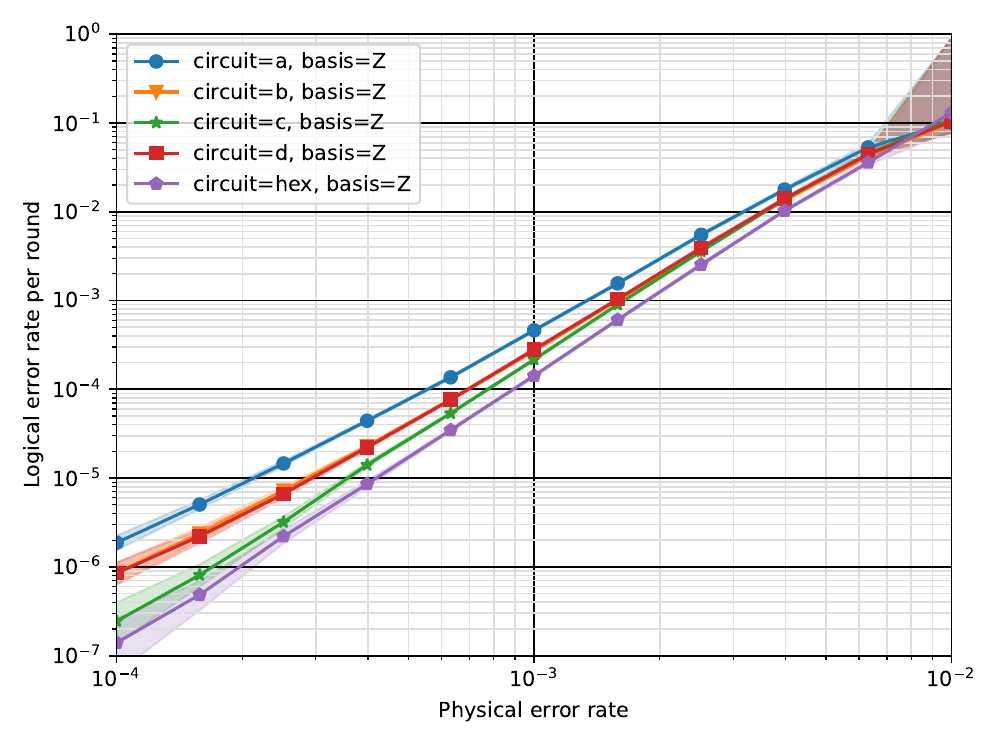}
    \caption{Physical error rate vs logical error rate per round using two-pass correlated sparse blossom for the four examples shown in \Cref{fig:lost_hex_component_examples} (a: broken qubit; b,c,d broken couplers) as well as for an unbroken hex-grid (labeled ``circuit=hex''). All circuits simulated use an SI1000 noise model and are 20-cycle memory experiments in the $X$ basis (top) and $Z$ basis (bottom).}
    \label{fig:benchmarking}
\end{figure}

In \Cref{fig:benchmarking} we present simulations of the example dropout configurations shown in \Cref{fig:lost_hex_component_examples}.
We use an SI1000 circuit-level noise model~\cite{Gidney2022benchmarkingplanar} simulated using stim~\cite{Gidney2021stimfaststabilizer} and a two-pass correlated sparse blossom decoder~\cite{Higgott2025sparseblossom,fowler2013optimal}.

For all four examples, each of which includes a broken qubit or coupler, and for most physical error rates considered, we find only a small increase in logical error rate of less than an order of magnitude relative to an unbroken hex-grid.
This is in contrast to the original version of LUCI introduced in \cite{debroy_luci_2024}, which would be unable to produce a valid circuit for cases (a), (c) or (d).
As expected, example (d) performs comparatively better for the $X$ memory experiment (since it preserves the $Z$ distance) and example (c) performs comparatively better in the $Z$ memory experiment (since it preserves $X$ distance).
Our simulations demonstrate that we can obtain good performance with hex-grid surface code circuits even in the presence of dropout.

\section{Conclusion and Outlook}\label{sec:outlook}
Hex-grid surface code circuits significantly relax the hardware requirements for quantum error correction, requiring far fewer components for a given distance and sparser degree-three connectivity.
However, the incompatibility of these circuits with existing methods for handling dropout was a major roadblock for their practical implementation, since many experimental quantum computing platforms inevitably suffer from the presence of fabrication defects.

In this work, we have proposed a new scheme for handling fabrication defects in hex-grid surface code circuits, extending the LUCI framework~\cite{debroy_luci_2024}.
Our approach results in the circuit distance reducing by one in the presence of an isolated broken qubit, as well as by one (in either basis or both) for an isolated broken coupler.
Through numerical simulations with a circuit-level noise model, we have shown that our method obtains good logical error rate performance in the presence of a broken qubit or coupler.

Our methods can also be applied directly to improve dropout handling in regular (four-coupler) surface code circuits.
For example, using the original LUCI method, a data qubit connected to two broken couplers would result in the distance reducing by one in both bases, whereas our approach reduces the distance by one in only one basis.
We do not expect the improvement to be as dramatic in this context, since the cases where we get an improvement for the four-coupler surface code are $O(p_{drop}^2)$ events, whereas the improvement applied to $O(p_{drop})$ events for hex-grid circuits (here $p_{drop}$ is the probability of a broken qubit or coupler).
In future work it would be interesting to study numerically the improvements obtained with our method for both square-grid and hex-grid surface codes for a realistic stochastic model of fabrication defects, and in a setting where some small fraction of the worst-performing devices can be removed from the sample.

By demonstrating a viable strategy for handling fabrication defects, we have removed a major barrier to the practical realization of surface codes in a hexagonal grid architecture.
In conjunction with the simpler hardware requirements and sparser connectivity of hex-grid surface code circuits, and therefore their improved scalability, our results further motivate the use of hex-grid architectures for realizing practical fault-tolerant quantum computation.
  
\section{Acknowledgements}
We thank Alexis Morvan and Adam Zalcman for helpful review of the manuscript. 
We would also like to thank the entire Google Quantum AI team for making an environment where this work is possible.

\bibliographystyle{alpha}
\bibliography{references}
\clearpage
\onecolumngrid

\appendix
\section{Additional LUCI diagrams}

\begin{figure*}[ht!]
    \centering
    \includegraphics[width=0.7\linewidth]{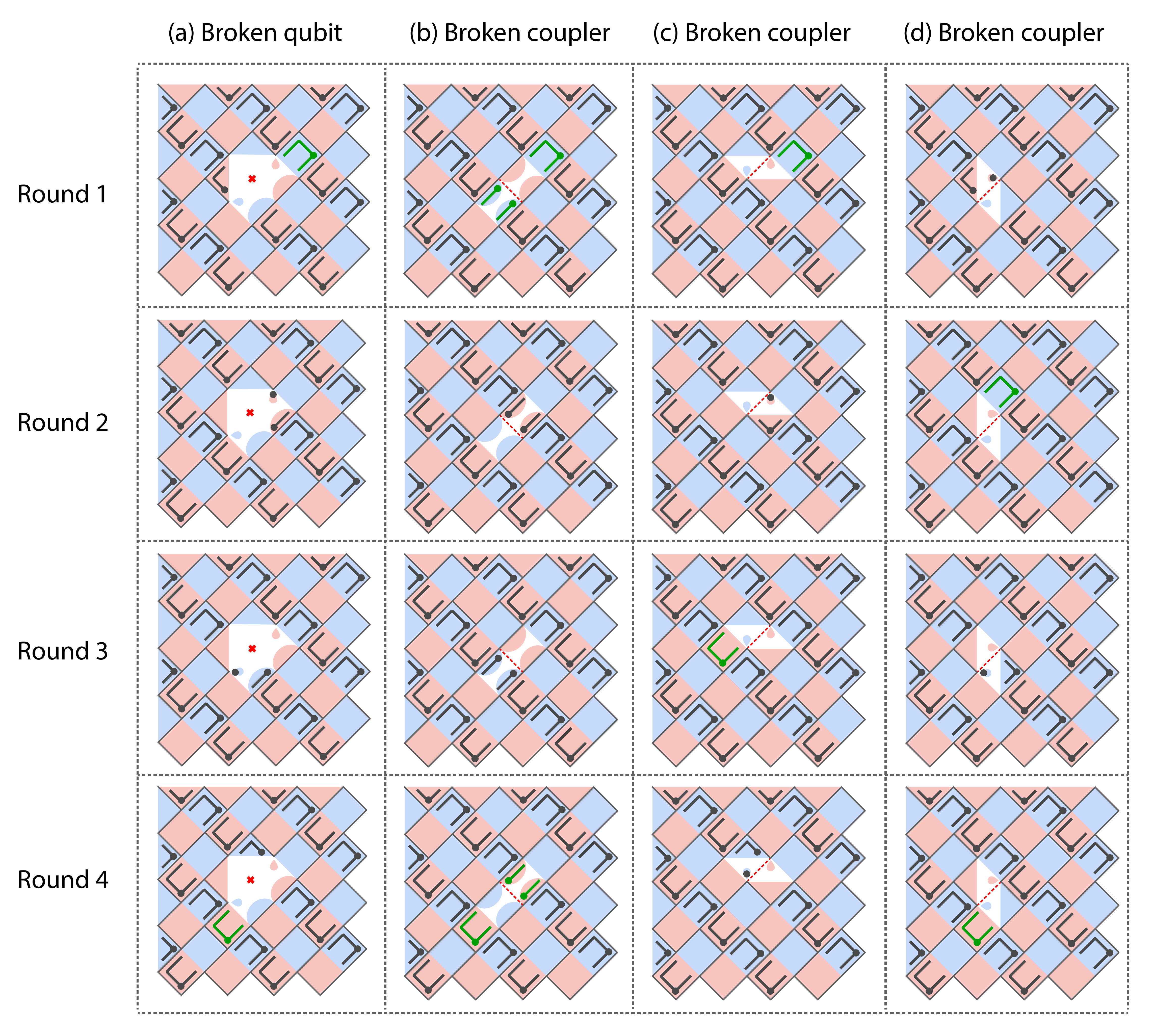}
    \caption{LUCI diagrams defining the circuits for all four cases from \Cref{fig:lost_hex_component_examples}. In the original version of LUCI, each gauge operator is measured once every four rounds. However, we introduce additional gauge operator measurements here (drawn in green) which introduce additional detectors. Links for opening the circuits in crumble: 
    \href{\caseabrokenqubithex}{Case A, Broken Qubit. }
    \href{\casebbrokencoupler}{Case B, Broken Coupler. }
    \href{\casecbrokencoupler}{Case C, Broken Coupler. }
    \href{\casecbrokencoupler}{Case D, Broken Coupler. }
    }
    \label{fig:placeholder}
\end{figure*}

\end{document}